# Hierarchical Testing of a Hybrid Machine Learning-Physics Global Atmosphere Model

Ziming Chen[1*], L. Ruby Leung[1*], Wenyu Zhou[1], Jian Lu[2,3], Sandro W. Lubis[1], Ye Liu[1], Chuan-Chieh Chang[1], Bryce E. Harrop[1], Ya Wang[4], Mingshi Yang[5], Gan Zhang[5], Yun Qian[1]

[1]Atmospheric, Climate, & Earth Sciences (ACES) Division, Pacific Northwest National Laboratory, Richland, Washington, USA

[2] College of Oceanic and Atmospheric Sciences, Ocean University of China, Qingdao, China

[3] State Key Laboratory of Physical Oceanography, Ocean University of China, Qingdao, China

[4] State Key Laboratory of Numerical Modeling for Atmospheric Sciences and Geophysical Fluid Dynamics, Institute of Atmospheric Physics, Chinese Academy of Sciences, Beijing, China

[5] Department of Climate, Meteorology, and Atmospheric Sciences, University of Illinois Urbana-Champaign, 1301 W. Green Street, Urbana, IL 61801, United States of America

**Key Points:**

- NeuralGCM, an ML-based hybrid model, shows comparable skill to physics-based models in simulating synoptic-scale extratropical cyclones.
- NeuralGCM captures the ENSO-induced responses and partially reproduces the nonlinear characteristics, albeit with overestimation.
- Under uniform warming, NeuralGCM simulates global-average responses and main tropospheric features similar to physics-based models.

*Corresponding author address: Ziming Chen and L. Ruby Leung, Atmospheric, Climate, & Earth Sciences (ACES) Division, Pacific Northwest National Laboratory, 902 Battelle Blvd, Richland, WA 99354. E-mail: ziming.chen@pnnl.gov, ruby.leung@pnnl.gov

**Abstract**

Machine learning (ML)-based models have demonstrated high skill and computational efficiency, often outperforming conventional physics-based models in weather and subseasonal predictions. While prior studies have assessed their fidelity in capturing synoptic-scale atmospheric dynamics, their performance across timescales and under out-of-distribution forcing, such as +3K or +4K uniform-warming forcings, and the sources of biases remain elusive, to establish the model's reliability for Earth science. Here, we design three sets of experiments targeting synoptic-scale phenomena, interannual variability, and out-of-distribution uniform-warming forcings. We evaluate the Neural General Circulation Model (NeuralGCM), a hybrid model integrating a dynamical core with ML-based component, against observations and physics-based Earth system models (ESMs). At the synoptic scale, NeuralGCM captures the evolution and propagation of extratropical cyclones with performance comparable to ESMs. At the interannual scale, when forced by El Niño-Southern Oscillation sea surface temperature (SST) anomalies, NeuralGCM successfully reproduces associated teleconnection patterns but exhibits deficiencies in capturing nonlinear response. Under out-of-distribution uniform-warming forcings, NeuralGCM simulates similar responses in global-average temperature and precipitation and reproduces large-scale tropospheric circulation features similar to those in ESMs. Notable weaknesses include overestimating the tracks and spatial extent of extratropical cyclones, biases in the teleconnected wave train triggered by tropical SST anomalies, and differences in upper-level warming and stratospheric circulation responses to SST warming compared to physics-based ESMs. The causes of these weaknesses were explored. Despite the noted weaknesses, NeuralGCM reproduces responses across experiments reasonably and performs comparably to ESMs. By integrating a dynamical core with ML, NeuralGCM shows potential for developing ML-based ESMs.

**Plain Language Summary**

Machine learning (ML)-based models hold great potential to transform how well we simulate weather and Earth's climate, with several recent successes. However, for these models to be trusted in Earth system science, they must produce simulations consistent with physical laws, even under conditions they have not encountered before. While ML-based models have been tested for weather forecasting, it remains uncertain whether they can produce reasonable responses in long-term simulations under unusual or novel forcings. A broad evaluation across different timescales is therefore essential. Besides, how well the emergent ML techniques can complement conventional physics-based models is still an open question. In this study, we present a series of idealized tests that cover systems at the synoptic scale, interannual scale, and under long-term out-of-distribution forcings. We test NeuralGCM, a hybrid model that combines a traditional differentiable solver for atmospheric dynamics with ML components, using a set of idealized experiments. NeuralGCM produces reasonable responses in all cases and performs similarly to conventional physics-based Earth system models, though with some limitations in simulating extratropical cyclone strength, atmospheric wave responses, and stratospheric warming and circulation responses. Overall, combining ML with established physics-based frameworks represents a promising path toward developing ML-based Earth system models.

1. **Introduction**

Earth system models (ESMs) are essential tools for understanding past climates and predicting possible future changes under illustrative emission scenarios (Flato et al., 2013; Zhou et al., 2020; IPCC, 2021; Ullrich et al., 2025). However, conventional physics-based ESMs remain computationally expensive and rely on simplified parameterizations of fundamental physical processes (Acosta et al., 2024; Eyring et al., 2024a; Ullrich et al., 2025). These limitations, combined with incomplete understanding of the Earth system, lead to persistent biases in simulating temperature (e.g., Flato et al., 2013; Meehl et al., 2020; Zelinka et al., 2020; Tebaldi et al., 2021), precipitation (e.g., Christopoulos & Schneider, 2021; Chen et al., 2021; Zhou et al., 2022; Chen et al., 2024), sea surface temperature (e.g., Li & Xie, 2012; Wang et al., 2014; Zhang et al., 2023), jet streams and blocking (e.g., Pithan et al., 2016; Woollings et al., 2018; Chemke et al., 2022; Chemke & Coumou, 2024; Tang et al., 2024), and tropical atmospheric circulation (e.g., Adam et al., 2016; Tian & Dong, 2020; Zhou et al., 2022; Chemke & Yuval, 2023; Ren & Zhou, 2024; Chen et al., 2024). Improving both the efficiency and reliability of ESMs remains a central research priority.

Recent advances in machine learning (ML) have led to the development of fast and accurate data-driven models for weather prediction (Eyring et al., 2024a, 2024b; Camps-Valls et al., 2025;

Ullrich et al., 2025). ML-based models trained on observations and reanalysis datasets now rival physics-based models in predictive skill, while offering substantial gains in computational efficiency (Keisler, 2022; Bi et al., 2023; Lam et al., 2023; Bouallègue et al., 2024; Rasp et al., 2024). They have demonstrated strong performance for both deterministic and ensemble forecasts across a wide range of spatial and temporal scales (Bi et al., 2023; Bouallègue et al., 2024; Kochkov et al., 2024; Pathak et al., 2024; Price et al., 2024). Recent studies mainly focused on synoptic-scale forecasting (Bouallègue et al., 2024; Rasp et al., 2024; Zhou et al., 2025; Husain et al., 2025; Sun et al., 2025), with limited efforts devoted to predictions at subseasonal (Diao & Barnes, 2025; Peings et al., 2025), seasonal (Zhang et al., 2025; Kent et al., 2025) or even longer timescales (Zhang & Merlis, 2025; Baxter et al., 2025). Despite these advances, building trust in ML-based ESMs remains a critical challenge (Eyring et al., 2024b; Ullrich et al., 2025).

A key open question is whether ML-based models truly learn physical principles or simply reproduce statistical patterns from the training data. Hakim & Masanam (2024) addressed this by designing four idealized tests under realistic topography and climatological mean conditions. These tests included a steady tropical heating test with Matsuno-Gill response, an extratropical cyclone test, a geostrophic adjustment test, and an Atlantic tropical cyclogenesis test. Results from the ML-based global weather-forecasting model, Pangu-Weather (Bi et al., 2023), suggested that it encoded physically meaningful behavior consistent with theory (Hakim & Masanam, 2024). Other studies have evaluated specific cases such as extratropical cyclones and air-sea interactions (Bonavita, 2024; Wang et al., 2024; Baño-Medina et al., 2025), though challenges persist in representing geostrophic balance and the inner-core structure of tropical cyclones in current ML-based models, such as Pangu, FourCastNet and GraphCast (Bonavita, 2024).

Beyond short-term weather prediction, far less is known about how ML-based models perform in climate simulations or under out-of-distribution forcings. Prior idealized tests have focused on synoptic-scale processes (Bonavita, 2024; Wang et al., 2024; Hakim & Masanam, 2024; Baño-Medina et al., 2025). A recent work by Kent et al. (2025) extended evaluation timescales by using the ML-based weather model, ACE2, for seasonal prediction (Watt-Meyer et al., 2023). When initialized in November with persisted SST-anomaly forcings, ACE2 skillfully reproduced ENSO teleconnection patterns in the following winter, though with potential limitations in capturing extreme seasonal states extending outside the training datasets (Kent et al., 2025). Whether ML-based models can remain stable and physically credible in long-term climate simulations under sustained out-of-distribution forcings remains an open question (Eyring et al., 2024b). Besides, most previous studies have evaluated fully ML-based models that emulate physics-based models, leaving the role and limitations of ML components within hybrid physics-ML systems largely underexplored. Addressing these gaps will shed light on the development of next-generation ML-based ESMs (Kochkov et al., 2024; Camps-Valls et al., 2025; Ullrich et al., 2025).

Recently a hybrid model, Neural General Circulation Model (NeuralGCM), that combines a conventional dynamical core with an ML-based component, might provide possible pathway to simulate responses under the out-of-distribution forcings (Kochkov et al., 2024; Yuval et al., 2026). Despite potential risks of long-term instability, NeuralGCM has demonstrated strong skill in weather forecasting, seasonal prediction of tropical cyclones (Kochkov et al., 2024; Zhang et al., 2025), and capturing key dynamical processes such as geostrophic balance, tropical waves propagation, and extratropical eddy-mean flow interaction (Kochkov et al., 2024; Baxter et al., 2025). Recent studies have begun to explore NeuralGCM's applicability under warming climates. Duan et al. (2025) used NeuralGCM to simulate a historical heatwave event under SSP3-7.0 forcing scenario for the year 2050, representative of ~1.4 (0.9~2.3) K global warming relative to 1995–2014 (Lee et al., 2021). Similarly, Jiménez-Esteve et al. (2025) performed attribution analyses of four heatwave events after removing an estimated ~1.3K anthropogenic warming signal. Although these perturbations may still lie within the effective training distribution, they suggest that NeuralGCM has potential for simulating warming-related responses. Kochov et al. (2024) previously examined NeuralGCM's generalizability in uniform SST-warming experiments, showing that the model reproduced several robust features associated with moderate +1K and +2K SST increases, but diverged from physics-based expectations under substantial SST increases (+4K). Zhang & Merlis (2025) also conducted +2K SST uniform-warming experiment and emphasized the role of the dynamical core in constraining NeuralGCM's warming responses based on detailed analysis. Despite that, earlier uniform-warming studies relied on single realizations, whereas robust characterization of forced responses typically requires ensemble simulations. Additionally, NeuralGCM's behavior under persistent out-of-distribution forcings and the mechanism governing its successes and failures remain largely unexplored.

To address these gaps, we design three idealized experiments to evaluate ML-based models across multiple timescales: (1) synoptic-scale weather systems, (2) atmospheric responses to interannual SST variability, and (3) responses to out-of-distribution uniform-warming SST forcings. We address two key questions: 1) Does NeuralGCM encode physically meaningful atmospheric dynamics? 2) If so, can it generate credible responses to interannual SST variability and to sustained out-of-distribution warming forcings? Here we use the stochastic version of NeuralGCM at 2.8º horizontal resolution, with 37 vertical pressure levels and 1-hour time step.

The remainder of this paper is organized as follows. Section 2 describes the experimental designs and datasets. Section 3 presents results comparing NeuralGCM to observations and physics-based ESMs. Section 4 provides a summary and discussion.

## 2. Datasets and Methods
### 2.1 Observational and Reanalysis Datasets

We employ several observational datasets both to provide model inputs and to evaluate model performance. To identify ENSO years and reduce observational uncertainty and dependence, we use two monthly gridded observational SST datasets: 1) the Hadley Centre Global Sea Ice and Sea Surface Temperature version 1.1 (HadISST v1.1) from 1870 to present (Rayner et al., 2003) and 2) Extended Reconstructed Sea Surface Temperatures Version 5 (ERSST v5) from 1854 to present (Huang et al., 2017). For precipitation, we use the Global Precipitation Climatology Project version 2.2 (GPCP v2.2) spanning from 1979 to the present (Adler et al., 2003). To evaluate large-scale circulation patterns, we use ERA5, the fifth-generation global reanalysis produced by the European Centre for Medium-Range Weather Forecasts, which originally has a spatial resolution of 0.25º× 0.25º and 137 vertical levels from the surface to 80 km, spanning from 1950 to the present (Hersbach et al., 2019).

Because GPCP does not extend before 1979, ERA5 precipitation data are used as a substitute to evaluate anomalies during three pre-1979 ENSO events (1972/73, 1973/74, 1975/76) (Jiménez-Esteve & Domeisen, 2019).

## 2.2 Physics-based Model Simulations

We analyze daily and monthly outputs of air temperature, surface temperature, geopotential height, precipitation, vertical velocity, and horizontal winds from the first available realization of the atmosphere-only Atmospheric Model Intercomparison Project (AMIP) runs from 18 CMIP6 models (Table S1) (Eyring et al., 2016). To estimate uniform-warming responses, we further use AMIP-P4K simulations from seven CMIP6 models (Webb et al., 2017). All datasets span the period 1979–2014. Prior to computing multi-model means, all model simulation outputs are regridded to 2.5º×2.5º resolution, using bilinear interpolation for circulation patterns, and first-order conservative interpolation for temperature, geopotential height and precipitation (Seneviratne et al., 2021). For cyclone tracking and computing pattern correlation coefficients and normalized root-mean-square errors, all fields at their native resolutions are regridded to 2.8º×2.8º.

## 2.3 Design of Idealized Experiments

NeuralGCM is a hybrid ML-based model that integrates a differentiable atmospheric dynamical core with ML-based components (Kochkov et al., 2024; Yuval et al., 2026). We use the stochastic version of NeuralGCM trained on ERA5 atmospheric fields and Integrated Multi-satellitE Retrievals for Global Precipitation Measurement (IMERG) version 7 precipitation from 2001 to 2018, with 2.8º spatial resolution and 37 vertical $\sigma$-levels (Yuval et al., 2026). NeuralGCM predicts the atmospheric states at the next time step using prescribed SST and sea-ice concentration, producing deterministic and ensemble forecasts with comparable or superior skill to both pure ML- and physics-based models (Kochkov et al., 2024). Compared to CMIP6 models – which typically have 1.4º×1.4º horizontal resolution and ~58 vertical levels – NeuralGCM operates at coarser spatial and vertical resolutions. Additional model details can be found in Kochkov et al. (2024) and Yuval et al. (2026).

We design three classes of idealized experiments targeting (1) synoptic-scale systems, (2) atmospheric responses to interannual SST variability, and (3) responses to long-term uniform SST warming, with the latter focusing on NeuralGCM's behavior under out-of-sample or out-of-distribution forcings (Fig. 1 and Supplementary Table S2). All experiments are initialized from ERA5 atmospheric states and run with hourly integration, using daily-updated ERA5 SST and sea-ice concentration. Following NeuralGCM documentation (https://neuralgcm.readthedocs.io/en/latest/checkpoint_modifications.html), we fix the global-mean surface pressure to enhance numerical stability. Model performance is evaluated using pattern correlation coefficients (PCCs) and normalized root-mean-square errors (NRMSEs) between NeuralGCM and observation/physics-based models. The NRMSEs have been normalized by the spatial standard deviation of the observations or physics-based models.

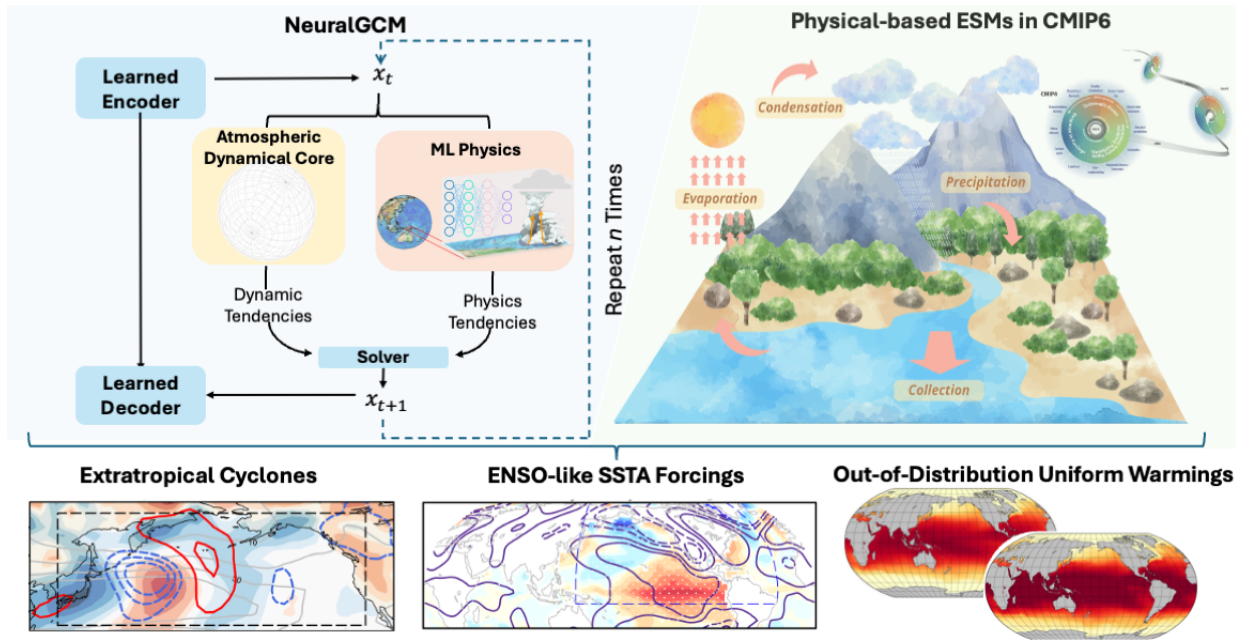

**Figure 1. Schematic diagram summarizing the hierarchical testing simulations conducted with NeuralGCM and CMIP6 ESMs.** The top panels illustrate the core structure of the NeuralGCM model and the setup of AMIP simulations in the CMIP6 ensemble. The bottom panels highlight the three testing simulations: extratropical cyclonic (ETC) and atmospheric response to ENSO-like SST anomaly (SSTA) and uniform SST warming. The left two bottom panels show the NeuralGCM-simulated responses in anomalous geopotential height and air temperature for the ETC and ENSO-like SSTA experiments, while the right bottom panel shows the setup of the uniform warming experiments.

### 2.3.1 Extratropical Cyclone (ETC) Experiments

ETCs are identified in ERA5 and CMIP6 models as local maxima in 1000-hPa geostrophic relative vorticity exceeding $10^{-4}$ s$^{-1}$ using 6-hourly data, following Hakim (2003). We focus on ETC wave packets over the western North Pacific (38ºN–42ºN, 148ºE–152ºE) during boreal winter (December–February, DJF) from 1979 to 2020. To avoid duplication, any cyclones forming within 24 hours and within the same domain (38ºN–42ºN, 148ºE–152ºE) after an initial detection are removed. This yields 526 ETC cases in the observations.

For each case, NeuralGCM is initialized with ERA5 atmospheric states on cyclone-formation date and integrated for four days, forced by the corresponding daily SST and sea-ice concentration prescribed to the NeuralGCM as boundary conditions. Skill is assessed through PCCs and NRMSEs of geopotential height and temperature over the North Pacific (30º–60ºN, 140ºE–120ºW). A parallel experiment is conducted for 453 summertime ETCs (June–August).

Since NeuralGCM ETC simulations are case-initialized with realistic ERA5 atmospheric states, whereas CMIP6 AMIP runs are continuous free-running simulations, NeuralGCM may benefit from the constraints of the initial conditions. To enable an "apples-to-apples" comparison, with the same configuration as the CMIP6 AMIP protocol, we also conduct long-term free-running AMIP transient simulations with NeuralGCM, driven by daily SST and sea ice from 1979 to 2014. Three ensemble members are generated using different initial conditions, hereafter referred to as "NeuralGCM AMIP runs" for convenience. When comparing NeuralGCM with CMIP6 AMIP runs, we use these free-running NeuralGCM AMIP simulations.

### 2.3.2 ENSO-like SST Anomalies (SSTA) Experiments

To assess the model's ability to simulate ENSO-induced anomalies, composites of large-scale anomalies during strong El Niño and La Niña years are computed using observations, the NeuralGCM AMIP runs, and the CMIP6 AMIP runs. El Niño (La Niña) events are defined as DJF-mean, detrended Niño3.4 SSTA exceeding $\pm 2$ standard deviations during 1950–2020 (Jiménez-Esteve & Domeisen, 2019). As NeuralGCM AMIP runs have three ensemble members, multi-member composites are used to obtain robust ENSO responses.

### 2.3.3 Uniform Warming Experiments

We design a set of long-term SST-uniform-warming runs from 1979 to 2014, following the experimental configuration of CMIP6 AMIP-P4K runs. The configurations mirror that of the NeuralGCM AMIP runs, but with SST uniformly warmed by +2K and +4K, respectively. Three ensemble members are produced (Supplementary Table S2). The average responses across the multi-uniform-warming scenarios, referred to as "AMIP-PxK" for convenience. These two sets of uniform-warming simulations are referred to as "36-year AMIP-PxK" for convenience.

To validate the uniform-warming responses in NeuralGCM under large-ensemble simulations, we further design both one-year AMIP and AMIP-PxK uniform-warming experiments with NeuralGCM. In the one-year AMIP configuration, SST and sea-ice concentration are prescribed as the 1979–2020 climatological mean. To reduce internal variability and yield a more robust response, we perform 20 ensemble members initialized on the first days of consecutive months from January 2018 to August 2019. Each simulation spans one year (Jan 1$^{st}$ to Dec 31$^{st}$) with SSTs uniformly increased by +1K, +2K, +3K, and +4K relative to the one-year AMIP baseline. Each warming scenario includes 20 ensemble members initialized identically to the one-year AMIP ensemble. Observed SST interannual variability rarely exceeds 2 K across most regions (Bulgin et al., 2020), so the +3K and +4K experiments represent out-of-distribution forcings.

It should be noted that stochastic version of NeuralGCM generates space-time correlated Gaussian random fields by random seeds, which both perturb the initial conditions and introduce stochasticity into the neural network parameterization (Kochkov et al., 2024; Zhang et al., 2025). In contrast, perturbation using different initial conditions alone allows the computation of ensemble means while primarily sampling initialization uncertainty (Brenowitz et al., 2024; Jiménez-Esteve et al., 2025). Therefore, in principle, both perturbation of random seeds and initial conditions are required to generate ensembles of statistically independent forecasts.

To facilitate comparison across different uniform-warming scenarios, responses are computed as the difference between AMIP-PxK and AMIP-like simulations, normalized by the global-mean SST warming magnitude. We mainly describe the results of the 36-year AMIP-PxK runs, while the responses of the one-year AMIP-PxK runs are presented in the Supplementary Information.

## 3. Results

### 3.1 NeuralGCM Captures the Evolution of Extratropical Cyclone but Overestimates Their Spatial Extent at Synoptic Scale

The western North Pacific is a canonical source region for ETC genesis, with systems typically developing and propagating downstream (e.g., Gyakum & Danielson, 2000; Hakim, 2003; Yoshida & Asuma, 2004). Our first experiment examines the time evolution of localized ETCs forming in this region. Figure 2 and Supplementary Figure S1 present the composites of anomalous geopotential height (zg500 & zg1000) and air temperature (ta500 & ta1000) at 500 and 1000 hPa, derived from 526 observed cases.

In observations, ETCs move eastward and slightly northward at 1000 hPa after formation (Supplementary Fig. S1 left). At upper levels, the trough develops into a wave packet that propagates along the westerly jet (Figs. 2A, 2C, 2E and 2G). By day 3, the system reaches the central North Pacific (Fig. 2G). Warm and cold anomalies appear ahead of and behind the ETC centers (gray contours), respectively, producing strong horizontal temperature gradient near the surface cold front that support trough intensification.

NeuralGCM simulations initialized with ERA5 atmospheric conditions reproduce the observed evolution of the cyclones and wave packet with reasonable fidelity. PCCs for zg500 exceed 0.8 and NRMSEs remain below 0.9 during the first four days, indicating skill in representing ETC dynamics. However, the dispersion characteristics in NeuralGCM is somewhat different, with the downstream ridges and troughs dissipating at a faster rate. Similar skill is found for zg1000 (Supplementary Fig. S1). The summer ETC experiments yield consistent but weaker anomalies, reflecting reduced cyclone development during the warm season (Supplementary Fig. S2).

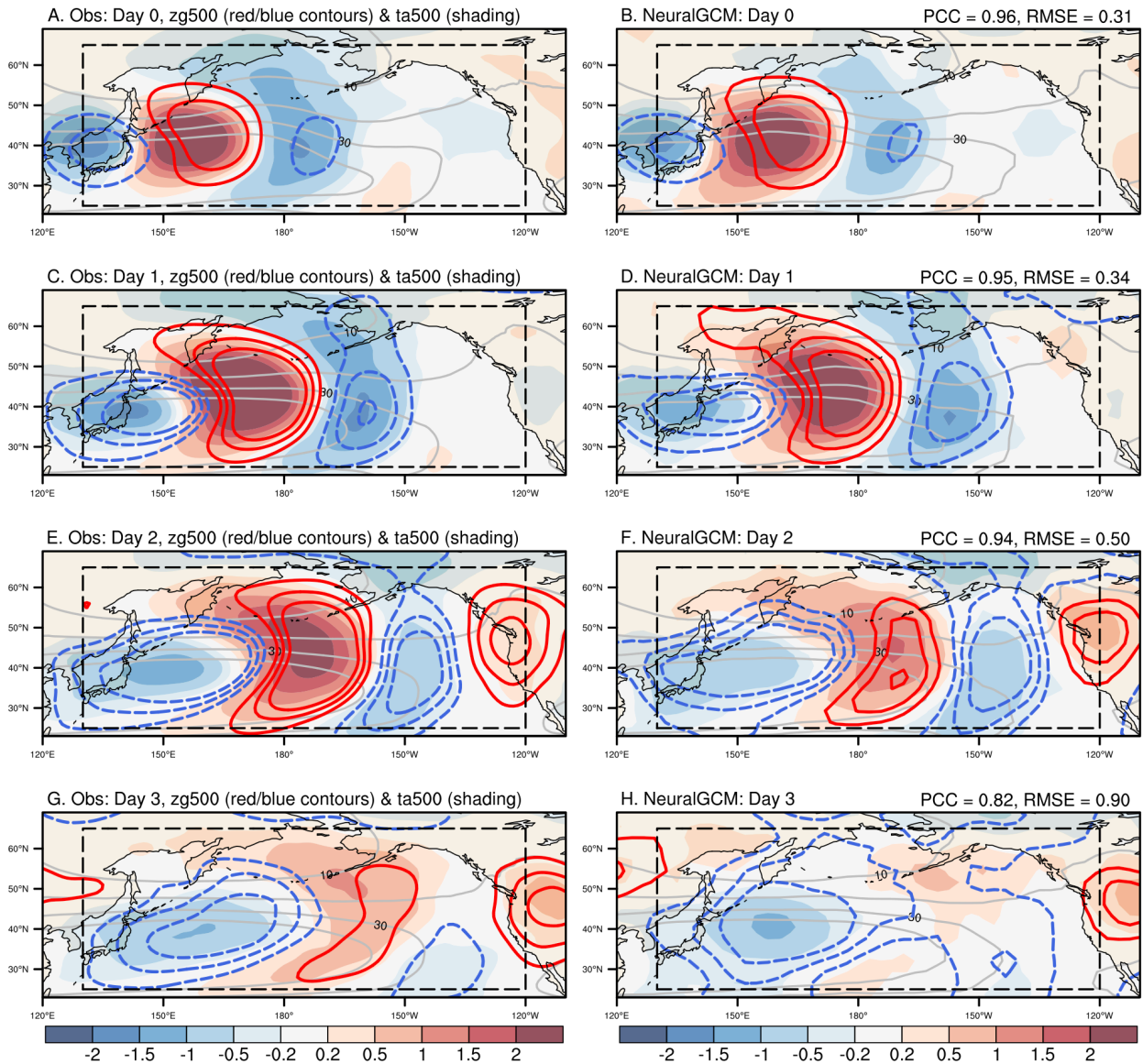

**Figure 2**. Composite of anomalous 500-hPa geopotential height (zg500, colored contours; units: m) and air temperature (ta500, shading; units: K) during boreal winter on (A, B) day 0, (C, D) day 1, (E, F) day 2, and (G, H) day 3 after the formation of extratropical cyclones (ETCs) in (A, C, E, G) observations and (B, D, F, H) NeuralGCM simulations. Anomalies are computed as deviations from the long-term monthly mean of ERA5. The composite is based on 526 ETC cases in observations. The quantities in the top-right corner represent the pattern correlation coefficients (PCCs) and normalized root-mean-square errors (RMSEs) of anomalous zg500 between observations and NeuralGCM simulations over the North Pacific (black dashed box). RMSEs were normalized by the spatial standard deviation over the North Pacific in observations (black dashed

boxes). Gray contours show the simultaneous composites of westerly wind at 200 hPa (ua200) in (A, C, E, G) ERA5 and (B, D, F, H) NeuralGCM simulations, with an interval of 10 m s$^{-1}$. The anomalous zg500 is shown by contours at intervals of 20 m in (A) and (B), 10 m in (C) and (D), 5 m from (E) to (H) starting from 10 m. The zero contour is not shown.

Despite these strengths, NeuralGCM tends to overestimate ETC spatial extent and exhibit track biases. By day 3, simulated systems become more expansive and intense than observed (Supplementary Figs. 1G, 1H). These errors likely arise from biases in storm position variability and storm-scale dynamics, which together amplify the spatial footprint of ETCs. Storm-centered composites (Supplementary Fig. S3) confirm that the magnitude and spatial scale of ETCs – as represented by negative zg1000 anomalies – are overestimated by 50%~63% on day 3. At mid-tropospheric levels, NeuralGCM produces negative zg anomalies of roughly –100 m over a region substantially broader than in observations. While notable biases emerge by day 3, NeuralGCM clearly captures the key physical processes governing ETC development through days 0–2.

Why does NeuralGCM perform well during the first two days but diverge by day 3? Upper-level circulation and potential vorticity (PV) provide important insight, as both strongly influence ETC development (e.g., Hakim, 2003; Heo et al., 2019; Kautz et al., 2022; Ni et al., 2025). Figure 3 shows composite PV and anomalous meridional wind (V300) at 300 hPa. In observations, upper-level PV anomalies align with anomalous V300 (contours in Figs. 3A, 3C, 3E, 3G). Mid-to-low-level convective centers (blue contours in Figs. 2, S1) propagate northeastward along the upper-level meridional PV gradient (shading in Fig. 3). NeuralGCM reproduces these PV and circulation patterns well through day 2 (Figs. 3A–3F). But NeuralGCM generally exaggerates the meridional PV gradient. By day 3, the simulated meridional PV gradient across the North Pacific is ~1.5 times stronger than observed, particularly between 180º~150ºW (Figs. 3G, 3H). This enhanced PV gradient supports stronger ascent and convection, driving the overly large ETC system across the basin (Figs. 2H & Supplementary Fig. S1H). Besides, the amplified PV gradient over central-to-eastern North Pacific also guides the ETC to propagate northeastward, leading to the biases in ETC tracks.

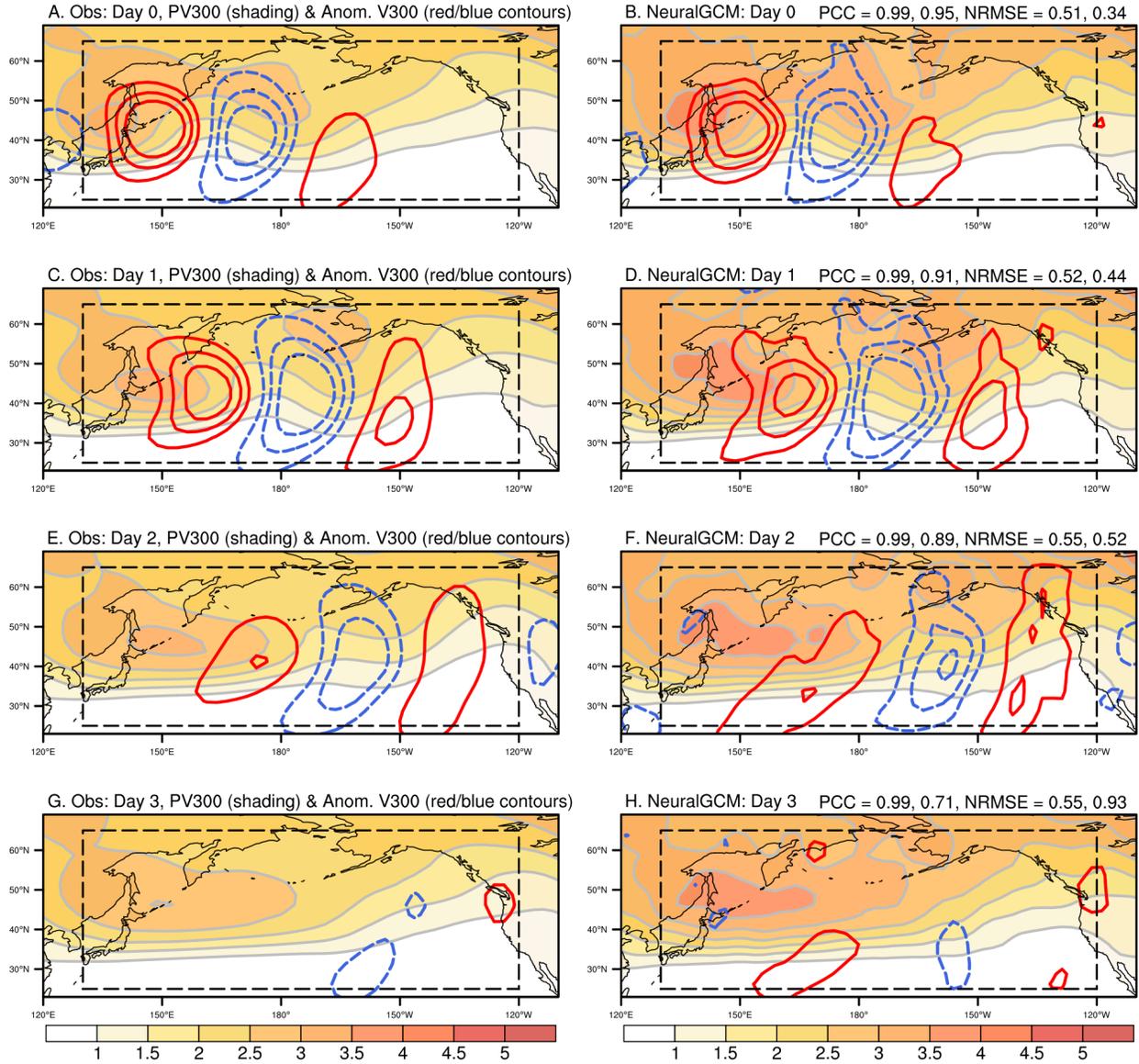

**Figure 3. Same as Figure 2, but for full-field potential vorticity (PV; shading; units: PVU = $10^{-6}$ m$^2$ K (kg s)$^{-1}$) and anomalous meridional wind (V300; units: m s$^{-1}$) at 300 hPa**. The anomalous V300 is shown by contours at intervals of every 2 m s$^{-1}$ starting from ±2 m s$^{-1}$. The zero contour is suppressed.

To benchmark NeuralGCM against physics-based ESMs, we compare PCCs and NRMSEs of anomalous zg1000, zg500, ta1000, and ta500 composites between models and observations. CMIP6 models conducted long-term free-running AMIP simulations which include mean state biases. Here to eliminate the advantages due to case initialization and to account for mean-state biases, these comparisons with physics-based ESMs use the long-term NeuralGCM AMIP runs driven by observed daily SST and sea-ice concentration. For zg1000, NeuralGCM achieves a higher PCC than eight selected CMIP6 models and lower NRMSE than three out of the eight models (Figs. 4A & 4B). Comparable performance is found for other variables (Figs. 4B–4D).

Since NeuralGCM was trained on the ERA5 and IMERG datasets for 2001–2018 (Yuval et al., 2026), the evaluations above include ETC cases that fall within the model's training period, during which NeuralGCM is optimized for these atmospheric states. To assess out-of-sample skill, we exclude all ETC cases within the training years. Supplementary Figure S4 shows the composites of anomalous zg500 and ta500 for ETCs during boreal winter in 1979–2000. NeuralGCM performs reasonably well in both the full period (1979–2020) and the out-of-sample period (1979–2000), but the performance is noticeably better when training-period events are included. For instance, the mean NRMSE of zg500 during days 1–3 is 0.38 for 1979–2020, approximately 18% lower than the value of 0.45 for 1979–2000. This reduction in the model performance can be attributed to the generalizability of the ML component, while the ETC simulation biases outside the training period originate from both the ML component and the dynamical core. Future work quantifying the relative contributions of these error sources could use a NeuralGCM configuration that allows the ML component to be selectively disabled for comparison with analytical solutions or other dynamical cores (including the NeuralGCM dynamical core at higher resolution) to isolate the errors associated with the dry dynamical core as applied to the low resolution in this study.

Overall, NeuralGCM implicitly encodes seasonally varying physical processes involved in ETC development. Its performance is broadly comparable to AMIP runs of physics-based ESMs, owing to the integration of a dynamical core with ML-based techniques. Nonetheless, the NeuralGCM systematically exaggerates the upper-level PV gradient, resulting in overly large ETC systems by day 3.

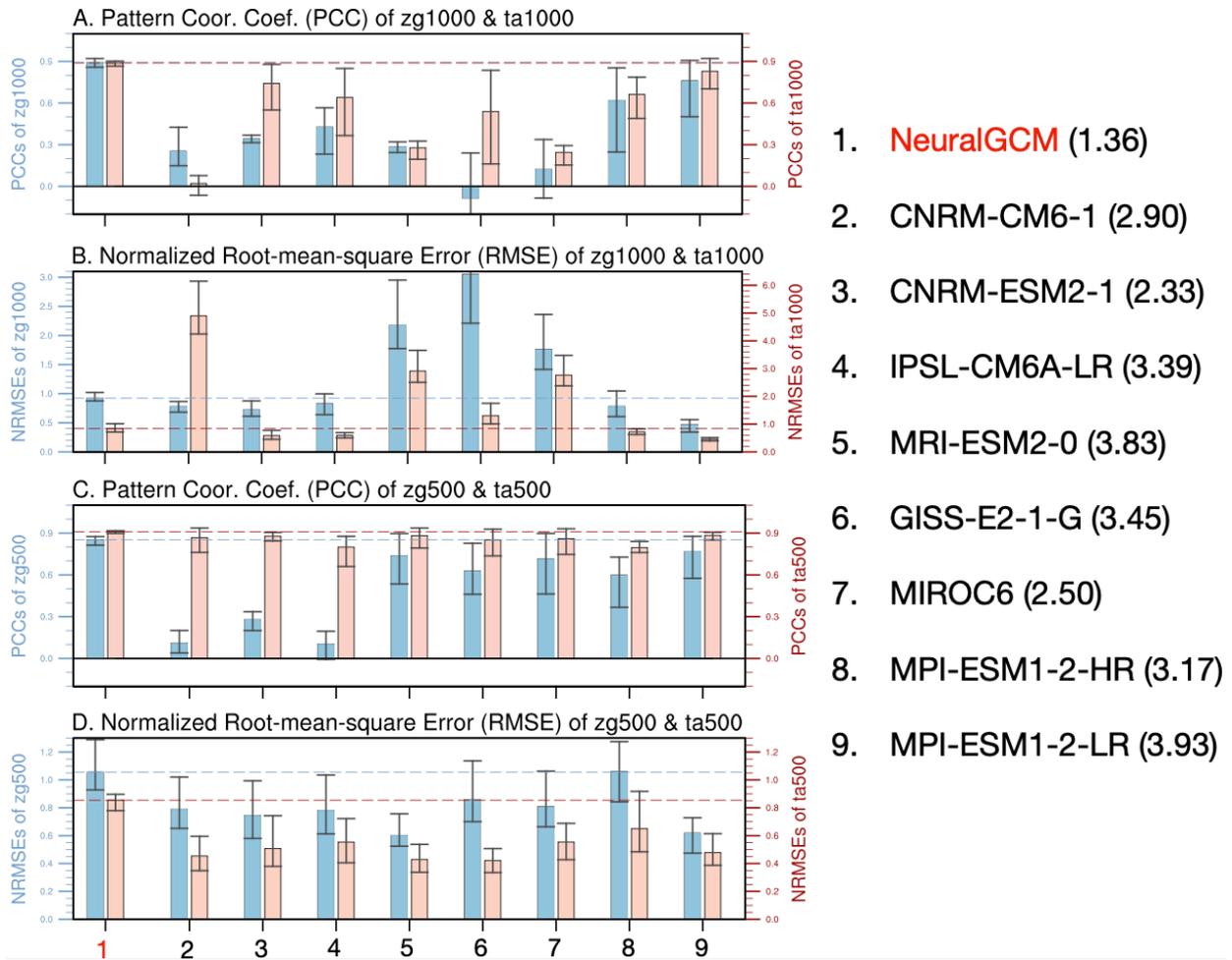

**Figure 4. Comparison of extratropical cyclone (ETC) simulations between the AMIP transient runs of NeuralGCM and physics-based models.** The ensemble mean across three members in NeuralGCM AMIP runs in 1979–2014 is shown. Bar charts in A and B show pattern correlation coefficients (PCCs) and normalized root-mean-square errors (NRMSEs) of anomalous geopotential height (blue) and air temperature (red) at 1000 hPa, averaged from day 1 to day 3 after cyclone formation, respectively, for models and observations over the North Pacific (30ºN–60ºN, 140ºE–120ºW). RMSEs were normalized by the observed spatial standard deviation. Vertical lines represent the range of PCCs and RMSEs across lead times of day 1 to day 3. Horizontal dashed blue and red lines indicate the performance of NeuralGCM in simulating geopotential height and air temperature, respectively. The numbers in parentheses indicate the number of monthly ETC cases in the simulations. Panels (C) and (D) are the same as (A) and (C) but at 500 hPa.

## 3.2 NeuralGCM Reproduces ENSO-Induced Teleconnections but Overestimates Their Intensity

ENSO exerts a strong remote influence on global climate and weather conditions. Here, we analyze the wintertime circulation response during ENSO years. During El Niño years, observations show a pair of upper-level anticyclonic circulation anomalies over the tropical eastern Pacific, consistent with the Matsuno-Gill framework (Matsuno, 1966; Gill, 1980). A Rossby wave train emanates from the warm SSTA and propagates into the Pacific-North America (PNA) region, producing an anticyclonic anomaly over the subtropical eastern Pacific, followed by cyclonic and anticyclonic anomalies over the northeastern Pacific and northeastern North America (Fig. 5A). The cyclonic anomaly over the northeastern Pacific reflects a deepened and eastward-shifted Aleutian Low under El Niño conditions.

Both NeuralGCM and the CMIP6 AMIP simulations reproduce the correct sign and general structure of this teleconnection, as reflected in anomalous 200-hPa geopotential height (zg200) (Figs. 5B–5C). We quantify performance using PCCs and NRMSEs for zg200 and ta1000 anomalies against observations (Figs. 6A–6C). NeuralGCM achieves PCC values of 0.86 (zg200) and 0.91 (ta1000), falling well within the CMIP6 AMIP ensemble spread (zg200: 0.76 [0.65–0.87]; ta1000: 0.90 [0.52–0.97]; Figs. 6A–6B).

Model skill also extends to hydroclimate responses. Figures 5D–5F show precipitation anomalies normalized by climatological precipitation to reduce sensitivity to mean-state biases. In the tropics, NeuralGCM captures the large-scale structure of El Niño precipitation anomalies but with notable regional biases, particularly an underestimation of the negative anomalies over the tropical western and central Pacific (Fig. 5E). In the extratropics, observations reveal a PNA-like precipitation pattern, with positive anomalies over the tropical eastern Pacific, followed by negative and positive anomalies over the North Pacific and western North America (Fig. 5D). Both NeuralGCM and CMIP6 AMIP simulations qualitatively reproduce this pattern (Figs. 5E–5F). Quantitatively, NeuralGCM achieves a PCC of 0.91 and an NRMSE of 0.49, while the CMIP6 AMIP runs exhibit much lower PCCs (0.84 [0.79–0.89]) and higher NRMSEs (0.89 [0.66–1.11]) (Fig. 6C). The underestimated dry anomalies over the tropical western Pacific likely stem from biases in the low climatological gross moist stability and critical convection threshold, as tropical precipitation is approximately proportional to the ratio between surface net radiative flux and gross moist stability (Zhang, 1993; Johnson & Xie, 2010; Jiménez-Esteve & Domeisen, 2019; Zhou et al., 2019). Similar results hold for La Niña-like SSTA forcing (Fig. S5).

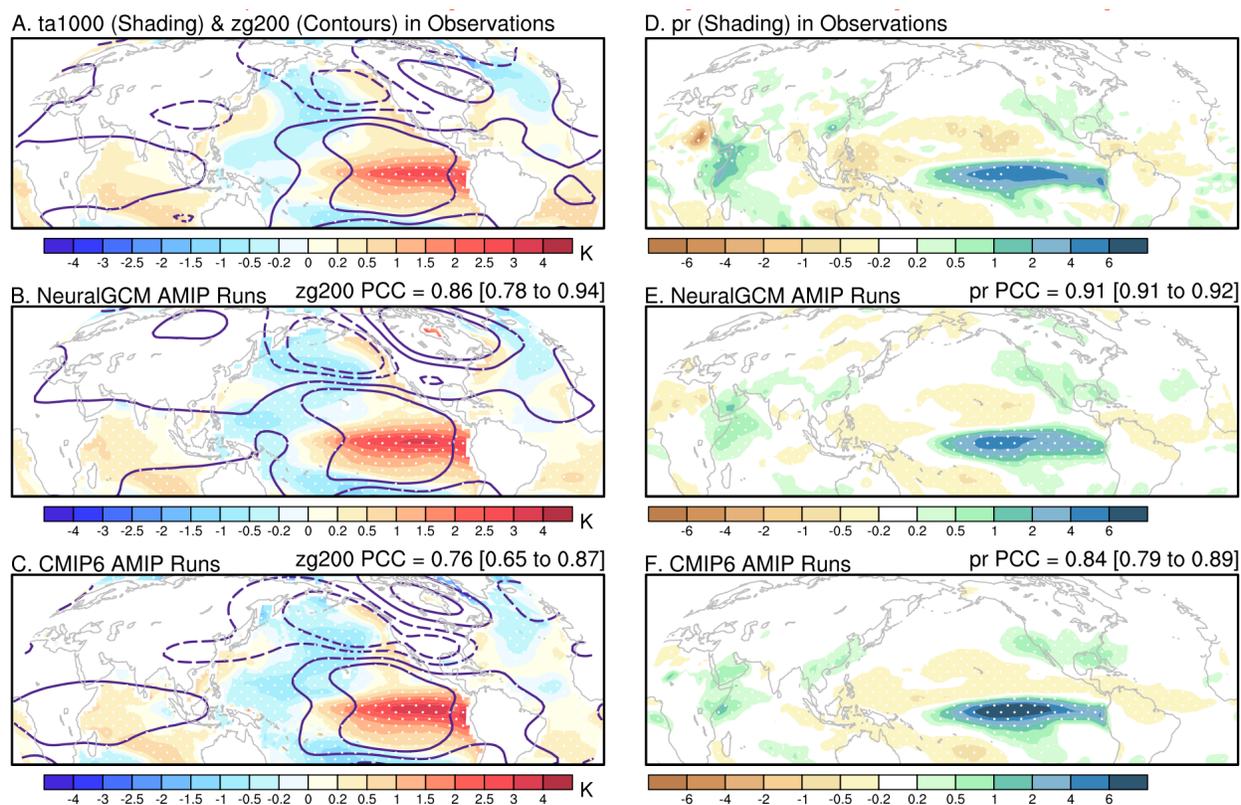

**Figure 5. Performance of NeuralGCM in simulating atmospheric responses to El Niño-like sea surface temperature (SST) anomalies and comparison against Earth system models.** (A) Composite of anomalous air temperature at 1000 hPa (ta1000, shading; units: K) and geopotential height at 200 hPa (zg200, contour; units: m; intervals: –50, –20, 20, 50 m) in El Niño years. Anomalies are defined as the deviation from the long-term mean from 1979 to 2020. (B) Multi-member composites of anomalous ta1000 and zg200 during El Niño years in NeuralGCM AMIP experiments in 1979–2014. (C) Same as (A), but for the multi-model mean composite during El Niño years in AMIP runs of seventeen CMIP6 models. White stippling in (A) and (B) represents significant anomalous ta1000 at the 10% level, while that in (C) indicates a consistent sign across >75% of CMIP6 models. Quantities in the top-right corner represent pattern correlation coefficients (PCCs) of zg200 between models and observations over the SH tropics and NH (10ºS–

90ºN, 0º–360º; blue dashed boxes). Multi-model means of PCC and NRMSE with their ranges are shown in (C). (D)–(F) are the same as (A)–(C), but for the composite of precipitation anomalies normalized by climatological precipitation (unitless).

Observational and modeling studies suggest that ENSO-induced teleconnections are nonlinear, due to nonlinear tropical convection response to underlying SSTA (e. g. Frauen et al., 2014; Zhang et al., 2014; Garfinkel et al., 2019; Jiménez-Esteve & Domeisen, 2019; Wang et al., 2023). We examine this nonlinear/asymmetrical response by evaluating the sum of the El Niño- and La Niña-induced response. In observations, besides the equatorial Pacific, strong asymmetry emerges in the northeast Pacific, where the PNA teleconnection is more pronounced during El Niño than La Niña (Fig. 6D).

To assess whether NeuralGCM captures this ENSO-induced asymmetry, we present the sum of its El Niño- and La Niña-year composites in the NeuralGCM AMIP simulations (Fig. 6E). Summing the composites isolates the nonlinear component associated with asymmetries in SST pattern and amplitude. NeuralGCM reproduces a PNA-like nonlinear pattern similar to observations and CMIP6 AMIP simulations (Figs. 6D & 6F). It captures the stronger PNA teleconnection during El Niño than during La Niña, reflected in a sequence of positive zg200 anomalies over the North Pacific, negative and positive zg200 anomalies across the northwestern and northeastern U.S. NeuralGCM also reproduces asymmetric tropical and extratropical precipitation responses across ENSO phases over the tropics and Northern hemisphere. The asymmetric pattern is broadly similar to observations, with a PCC of 0.78. However, NeuralGCM overestimates both the magnitude and spatial extent of the nonlinear component of zg200 and precipitation compared to observations and CMIP6 AMIP runs.

In summary, NeuralGCM successfully reproduces ENSO teleconnections and their associated hydroclimate responses, and partially captures their nonlinear characteristics, albeit with overestimation of teleconnection intensity.

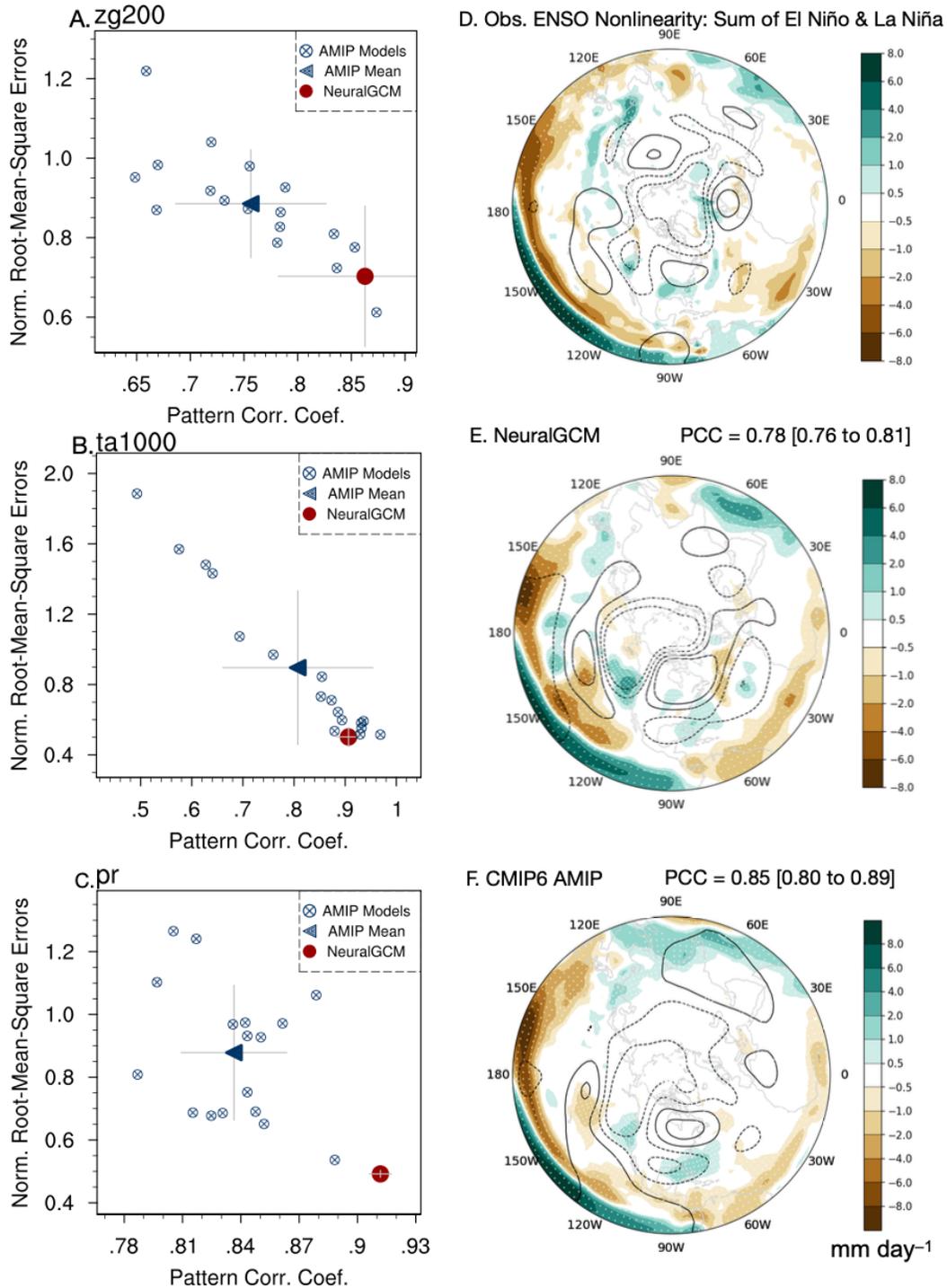

**Figure 6. Model performance of (A–C) El Niño-induce anomalies and (D–F) ENSO response asymmetry in boreal winter among NeuralGCM and CMIP6.** (A) Inter-model scatter plot between PCCs and normalized RMSEs of zg200 over the SH tropics and NH regions (10ºS~75ºN, 0º~360º). RMSEs were normalized by the spatial standard deviation of anomalies in the

observations. The red dot marks the performance of NeuralGCM with range of ±1 standard deviation across three ensemble members, while the triangle represents the multi-model mean with a range of ±1 standard deviation across CMIP6 models. (B) and (C) are the same as (A), but for ta1000 over the PNA region and precipitation, respectively. (D) Sum of anomalous composites during El Niño and La Niña years. Shading shows the asymmetrical response of precipitation (units: mm day$^{-1}$), while contours show the asymmetrical response of zg200 with intervals of 30 m. (E) and (F) are the same as (D), but for NeuralGCM and the multi-model mean of the CMIP6 AMIP runs. White stippling in (D) represents significant anomalous precipitation at the 10% level, while that in (E, F) indicates consistent sign across all NeuralGCM members or >75% of CMIP6 models, respectively. Quantities in the top-right corner represent pattern correlation coefficients (PCCs) of precipitation between models and observations over the SH tropics and NH (10ºS–90ºN, 0º–360º).

## 3.3 NeuralGCM Simulates Tropospheric Responses to Long-term Uniform Warming but Misrepresents Upper Troposphere-Lower Stratosphere Processes

Here, we evaluate NeuralGCM's responses to uniform SST warmings to assess its ability to generalize under out-of-distribution forcings. To verify whether NeuralGCM produces physically reasonable responses at the global scale, we compare globally averaged responses of ta1000 and precipitation between NeuralGCM and CMIP6 models relative to their AMIP runs (Figs. 7A–7B). Under the 4K uniform warming, the global mean ta1000 increases by 1.05 (1.01~1.09 for the mean ± one standard deviation across models) K K$^{-1}$ in CMIP6. NeuralGCM simulates comparable responses: 0.97 (0.90~1.04) K K$^{-1}$ and 0.75 (0.71~0.79) K K$^{-1}$ in 36-year AMIP-P2K and AMIP-P4K, respectively, and 1.01 (0.95~1.08) K K$^{-1}$, 0.94 (0.88~1.00) K K$^{-1}$, 0.89 (0.85~0.93) K K$^{-1}$, and 0.89 (0.87~0.92) K K$^{-1}$ in one-year AMIP-P1K, AMIP-P2K, AMIP-P3K, and AMIP-P4K, respectively (Fig. 7A). The average warming in ta1000 in 36-year AMIP-PxK is 0.86 (0.81~0.92) K K$^{-1}$, suggesting that NeuralGCM generally simulates weaker warming responses than those in CMIP6 models under SST-uniform-warming scenarios. Due to the lack of land feedbacks (Duan et al., 2025; Liang et al., 2025), it is expected that NeuralGCM underestimates the warming amplitude compared to physics-based models.

For precipitation, NeuralGCM's responses in AMIP-PxK are also comparable with the current physics-based ESMs. CMIP6 models show an increase of 3.44 (3.27~3.62) % K$^{-1}$ under 4 K warming. NeuralGCM simulates 4.20 (3.85~4.54) % K$^{-1}$ and 4.08 (3.94~4.22) % K$^{-1}$ in 36-year AMIP-P2K and AMIP-P4K runs, respectively, with a mean of 4.14 (3.90~4.38) % K$^{-1}$ (Fig. 7B). In the one-year uniform warming runs, NeuralGCM simulates 3.80 (3.46~4.15) % K$^{-1}$, 3.84 (3.63~4.05) % K$^{-1}$, 3.82 (3.70~3.94) % K$^{-1}$, and 3.68 (3.57~3.80) % K$^{-1}$ across AMIP-P1K to P4K, respectively. These results confirm that NeuralGCM captures the global mean responses generally consistent with physics-based ESMs, though with some discrepancies.

We next examine the zonal-mean vertical profile of the warming response (Figs. 7C–7D). In CMIP6 AMIP-P4K, the warming is characterized by pronounced upper-tropospheric amplification in the tropics (~2 K K$^{-1}$ between 30ºS~30ºN and 300 hPa~100 hPa), and muted warming in the extratropical upper troposphere and lower stratosphere (UTLS) poleward of 60ºN and above 250 hPa (Fig. 7C). The enhanced tropical upper-level warming is a robust response linked to moist-adiabatic lapse rate changes and increased latent heat release (Held, 1993; Wu et al., 2012). NeuralGCM broadly reproduces this tropospheric warming pattern (Fig. 7D), achieving a PCC of 0.62 between 1000 hPa and 100 hPa across 90ºS~90ºN. Despite this, significant discrepancies emerge in the UTLS of both hemispheres in NeuralGCM. In the tropical UTLS, NeuralGCM simulates excessive cooling, while in the extratropics, it overestimates warming above 250 hPa, compared to CMIP6 AMIP-P4K. Including the UTLS substantially reduces the PCC to 0.39 (Fig. 7D). A similar pattern appears in the one-year NeuralGCM AMIP-PxK simulations (Supplementary Figure S6B), although the UTLS response is somewhat weaker than that in the 36-year AMIP-PxK experiments. This stronger response may reflect the larger SST variability in the long-term integrations, because the one-year AMIP-PxK runs use climatological mean SST plus uniform warmings. As CO2 and ozone variations were not included as input during NeuralGCM's training, its stratospheric cooling in response to SST increases is likely a reflection of the observed anticorrelated relationship learned from ERA5 rather than a mechanistic response to $CO_2$ and ozone changes. Observational and modeling studies show that recent lower-stratospheric cooling is driven primarily by rising CO2 and declining ozone-depleting substances (Shine et al., 2003; Aquila et al., 2016; Mitchell, 2016). Thus, although NeuralGCM may have learned aspects of the historical tropospheric warming and stratospheric cooling from ERA5, its pronounced UTLS cooling under uniform SST warmings is likely nonphysical.

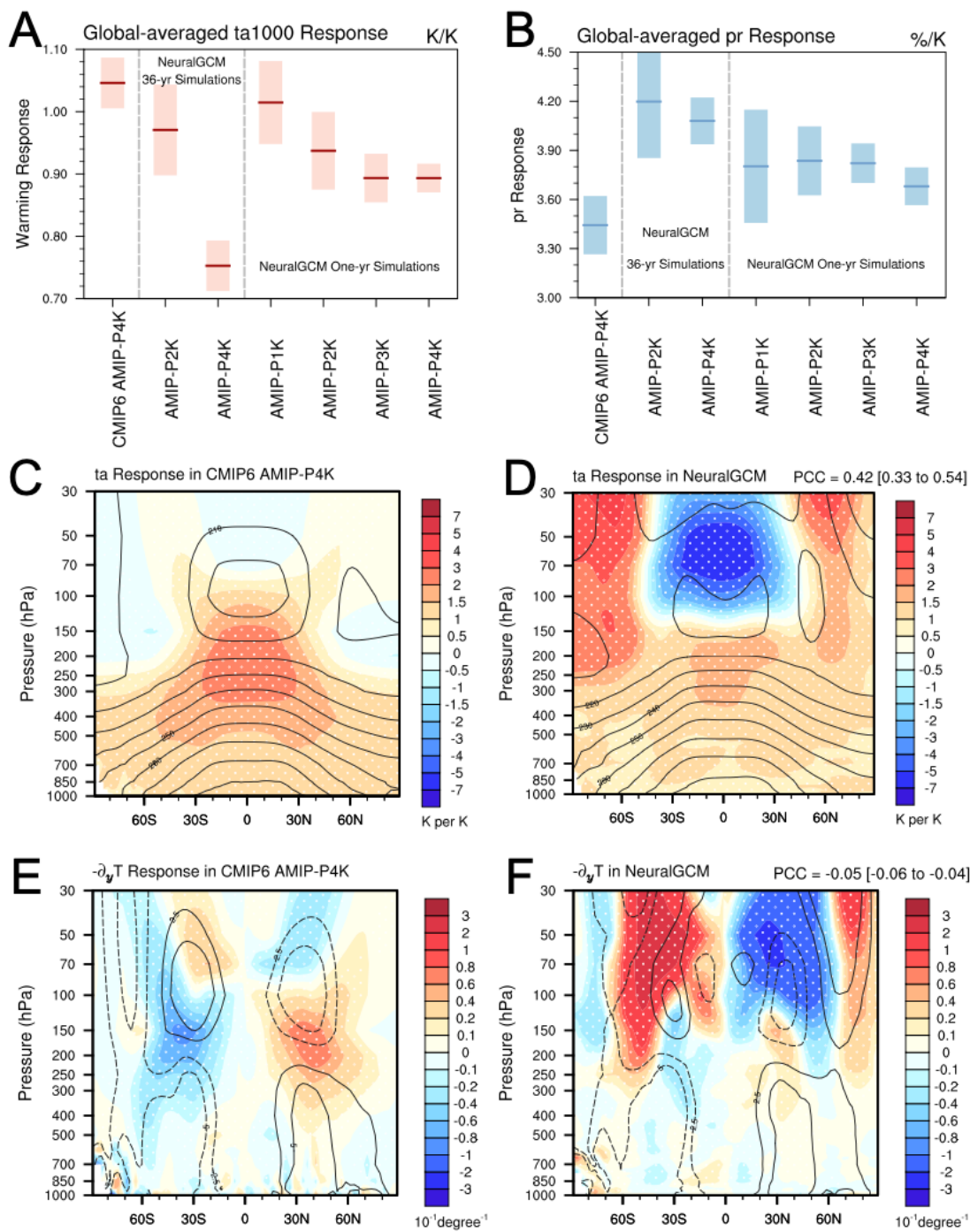

**Figure 7.** Uniform-warming response of air temperature (units: K K$^{-1}$) and precipitation (units: % K$^{-1}$) in NeuralGCM and CMIP6 models. (A) and (B) show globally-averaged responses of air

temperature at 1000 hPa and precipitation in CMIP6 AMIP-P4K (leftmost bar) and NeuralGCM 36-year (bars in central boxes) and one-year uniform warming runs (bars in right boxes), respectively. Thick horizontal lines represent multi-model or multi-member ensemble means, while bar charts show the range of ±1 standard deviation across models or members. (C) and (D) show the zonal-mean profiles of air temperature responses (shading) and present-day climatologies (contours) in CMIP6 AMIP-P4K and the average across NeuralGCM 36-year AMIP-P2K and AMIP-P4K runs, respectively (AMIP-PxK). White stippling in (C) denotes consistent sign of responses across >75% of the CMIP6 models, while that in (D) indicates significant responses at the 10% level across >50% of six ensemble members. (E) and (F) are same as (C) and (D) but for the zonal-mean temperature meridional gradients (Units: $10^{-1}$ degree$^{-1}$). The quantities on the top-right corner of (D, F) show the pattern correlation coefficients between CMIP6 AMIP-P4K runs and 36-year AMIP-PxK with range of ±1 standard deviation across six ensemble members.

These UTLS temperature biases directly affect the meridional temperature gradient ($-\partial_y T$) and the associated circulation responses (Figs. 7E–7F). Given that, we present the annual-mean responses in zonal-mean meridional temperature gradient ($-\partial_y T$) in CMIP6 AMIP-P4K and NeuralGCM 36-year AMIP-PxK (Figs. 7E & 7F). CMIP6 AMIP-P4K exhibits a weakening of $-\partial_y T$ in the midlatitude upper troposphere and a strengthening near the tropopause, consistent with robust warming patterns. NeuralGCM captures these features but exaggerates UTLS gradients and simulates spurious positive responses over the NH polar regions, reflecting its UTLS temperature biases.

Consequently, westerly jet responses differ between CMIP6 and NeuralGCM (Figs. 8A–8B). In CMIP6 AMIP-P4K, jets strengthen barotropically and shift poleward (Fig. 8A), driven by the strengthened meridional temperature gradient near the tropopause (Fig. 7E). These well-documented responses (e.g., Palipane et al., 2017; Lubis et al., 2018; Zhou et al., 2022) are quantified by the latitude shift of the peak zonal wind in the mid-troposphere (400–700 hPa). To measure the location changes in the westerly jet, we quantify the meridional jet shift based on the latitude difference of the peak zonal wind in the middle troposphere (400–700 hPa) between uniform-warming and AMIP runs, following (Zhou et al., 2022). The NH and SH jets show significantly poleward shifts by 0.38 (0.17~0.58) º K$^{-1}$ and –0.38 (–0.51~–0.24) º K$^{-1}$, respectively, in CMIP6 AMIP-P4K (Table S3). NeuralGCM reproduces the strengthening responses at and above the cores of the subtropical jets in the troposphere (Fig. 8B). It also captures the barotropic responses in westerlies and the poleward shift of NH westerlies. However, neither the NeuralGCM 36-year or one-year AMIP runs simulate the observed mean-state tropical stratospheric easterlies (contours in Figs. 8B & S6D). In the subtropical and extratropical UTLS, weakened meridional temperature gradients in NeuralGCM 36-year AMIP-PxK runs lead to negative zonal wind responses. Besides, NeuralGCM does not reproduce a statistically significant poleward shift of the westerly jet under uniform warming (Table S3). Averaged across the two uniform-warming runs, the NH westerly jet shifts are 0.20 (–0.17~0.58) º K$^{-1}$, which is insignificant at the 10% level. In contrast, the SH westerly jet is projected to shift equatorward by 0.64 (0.38~0.98) º K$^{-1}$.

Similarly, we present the responses of zonal-mean meridional streamfunction under uniform-warming scenarios (Figs. 8C & 8D). CMIP6 AMIP-P4K projects poleward expansion, with strengthened subtropical (~30º) and weakened tropical circulation (30ºS~30ºN; Fig. 8C), consistent with jet shifts (Fig. 8A). NeuralGCM reproduces subtropical expansions but fails to capture tropical weakening (Fig. 8D & Supplementary Fig. S6F), likely reflecting the inability of the data-driven convection scheme to representing the response to the out-of-distribution forcings. Therefore, NeuralGCM reproduces part of the responses of tropospheric large-scale circulation but fails to capture the circulation patterns in UTLS and tropical troposphere.

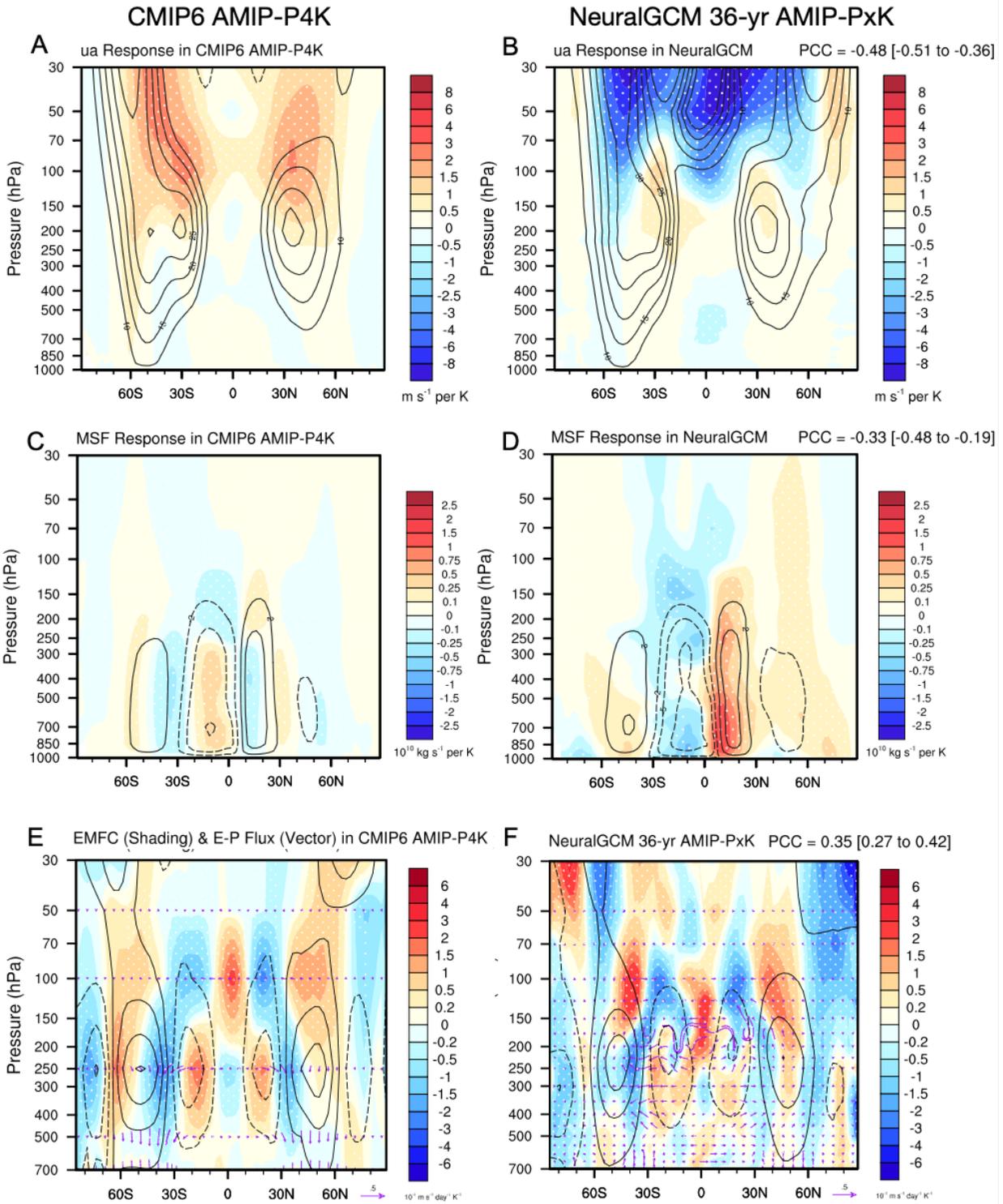

**Figure 8.** Comparison of NeuralGCM in simulating large-scale circulation responses to uniform warming against CMIP6 AMIP-P4K runs. In (A) and (B), shading shows annual uniform-warming responses in zonal-mean zonal wind (ua; units: m s$^{-1}$ K$^{-1}$) in CMIP6 AMIP-P4K and averaged

across NeuralGCM 36-year AMIP-P2K and AMIP-P4K runs, respectively. Contours represent present-day climatological ua in AMIP runs of CMIP6 and NeuralGCM. (C) to (D) are same as (A) and (B), but for the annual zonal-mean streamfunction (units: $10^{10}$ kg s$^{-1}$ per K). In (E) and (F), uniform-warming response (shading) of eddy momentum flux convergence (EMFC) is shown for CMIP6 AMIP-P4K and NeuralGCM averaged across 36-year AMIP-P2K and AMIP-P4K runs, respectively ($10^{12}$ m s$^{-2}$ per K). Contours represent present-day climatological EMFC in AMIP runs of CMIP6 and NeuralGCM. Stippling in (A, C, E) denotes sign agreement of >75% of the CMIP6 models, while stippling in (B, D, F) denotes significant responses in more than half of the six ensemble members. Vectors represent the response in Eliassen-Palm (EP) flux to the uniform warming. Purple vectors indicate significant responses in more than half of the models or members.

What factors contribute to the distinct responses between NeuralGCM and physics-based models? Given a close linkage between temperature and circulation responses in UTLS (Andrews et al., 1987; Butchart & Butchart, 2014; Lubis et al., 2016), we use the transformed Eulerian mean (TEM) framework (Test S2; Andrews et al., 1987; Rosenlof, 1995) to diagnose NeuralGCM's UTLS discrepancies. Figure S7 presents the potential temperature and streamfunction of residual circulation at the UTLS simulated by CMIP6 and NeuralGCM. Positive and negative values in streamfunction indicate clockwise and anti-clockwise circulation cells, respectively. Under the heating in the tropics, we could see an ascent and descent over equatorial and subtropical regions, shaping the air temperature patterns (Fig. S7A). But in the NeuralGCM, pronounced distinction can be found in the mean-state residual circulation. For example, over NH high latitudes (north of 60ºN), stronger descending motion than that in CMIP6 exists at UTLS in NeuralGCM (30 hPa ~ 100 hPa; Fig. S7B). The adiabatic heating leads to a warmer UTLS in the NH high latitudes (contours in Fig. 7D) than that in CMIP6 AMIP runs (contours in Fig. 7C). Under the uniform warming scenarios, potential temperature ($\bar{\theta}_t$) at UTLS largely contributes to the local warming responses, given similar patterns between $\bar{\theta}_t$ and temperature responses (shadings in Figs. 7C, 7D, S7C & S7D). Therefore, by diagnosing the $\bar{\theta}_t$, we can understand the sources of distinction in warming responses between CMIP6 AMIP-P4K and NeuralGCM 36-year AMIP-PxK. Compared to the weak warming responses in CMIP6 AMIP-P4K (Fig. 7C), cooling responses at tropical

UTLS in NeuralGCM 36-year AMIP-PxK (Fig. 7D) are caused by the pronounced responses in upward residual circulation which leads to strong adiabatic cooling (Fig. S7D). Reverse responses can be found in NH high latitudes (north of 50ºN).

To further explore the source of distinction in $\bar{\theta}_t$ responses between CMIP6 AMIP-P4K and NeuralGCM 36-year AMIP-PxK, we decompose the $\bar{\theta}_t$ responses into the contributions from diabatic ($\bar{Q}'$) and adiabatic heating ($\bar{Q}'_{adia}$) (Figure 9), using the TEM zonal-mean thermodynamic equation (Text S2). In CMIP6 AMIP-P4K runs, both diabatic and adiabatic heating contributions are small (Figs. 9A & 9C), leading to the weak overall $\bar{\theta}_t$ response. Despite that, the contributions from both terms in NeuralGCM AMIP-PxK runs are significantly larger than in CMIP6 AMIP-P4K (Figs. 9B & 9D). The $\bar{Q}'$ term only dominates the $\bar{\theta}_t$ responses over parts of equatorial and polar regions (Fig. 9B), while the $\bar{Q}'_{adia}$ term has main contributions to the pattern of $\bar{\theta}_t$ responses (Fig. 9D). Over NH and SH tropics, the negative contributions of $\bar{Q}'_{adia}$ shape the cooling responses in $\bar{\theta}_t$ and air temperature (Figs. 7D & S7D). Besides, over NH mid-low latitudes (20ºN~60ºN), positive $\bar{Q}'_{adia}$ contributes to local pronounced warming responses.

To investigate the dominant processes related to $\bar{Q}'_{adia}$, we further decompose it into vertical and horizontal advections due to residual circulation and eddy term (Eq. (S5) in Text S2). The vertical advection of potential temperature ($-\bar{\omega}^*\bar{\theta}_z$) dominates the responses in $\bar{Q}'_{adia}$ (Figs. 9E & 9F), while other terms only make secondary contributions (Fig. S8). Therefore, in the NeuralGCM uniform warming runs, the negative contributions of $\bar{Q}'_{adia}$ due to stronger responses in upward residual circulation and vertical advection in the deep tropics lead to a stronger cooling response than that in CMIP6 AMIP-P4K. In contrast, in NH mid latitudes (20ºN~60ºN) and SH polar regions (65ºS~90ºS), positive contributions of $\bar{Q}'_{adia}$ due to stronger responses in downward residual circulation led to larger warming responses than those in CMIP6 AMIP-P4K. Besides, the $\bar{Q}'$ term also makes important contributions to the temperature responses in equatorial and SH polar regions.

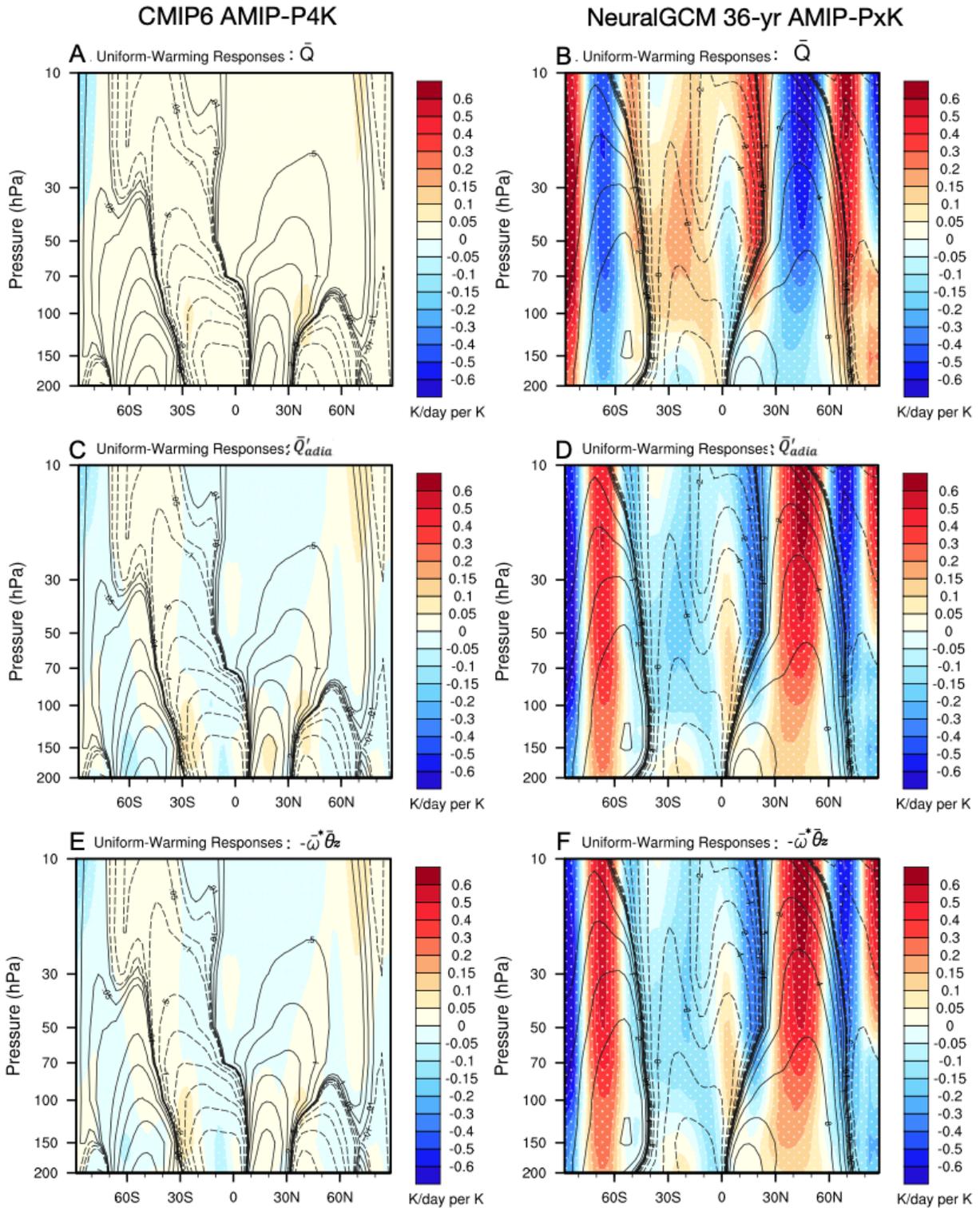

**Figure 9. Diagnostic analysis of zonal-mean Transformed Eulerian-Mean (TEM) thermodynamic equation.** (A, B) contributions of total diabatic heating ($\bar{Q}$) to the responses of potential temperature tendency ($\bar{\theta}_t$) in (A) CMIP6 AMIP-P4K and (B) the average across

NeuralGCM 36-year AMIP-P2K and AMIP-P4K runs (AMIP-PxK). Contours denote the responses in residual streamfucntion ($\psi_{tot}$; units: $10^8$ kg m$^{-1}$ s$^{-1}$ K$^{-1}$), with intervals of $\pm 0.01$, 0.05, 0.1, 0.5, 1, 2, 4, 8, 16, 32. (C) and (D) are same as (A) and (B) but for the contributions of total adiabatic heating ($\bar{Q}'_{adia}$). (E) and (F) are same as (A) and (B) but for the contributions of vertical advection for potential temperature due to residual circulation ($-\bar{\omega}^* \bar{\theta}_z$). Units: K day$^{-1}$ per K

How can we understand the differences in tropospheric circulation responses to uniform warmings between CMIP6 and NeuralGCM? Previous studies suggest that eddy momentum flux convergence (EMFC), associated with meridional propagation of wave activity, maintains the westerly jet at mid-high latitudes (Lu et al., 2015; Palipane et al., 2017). The poleward shift of the westerly jet and expansion of the Hadley cell under warming conditions have been attributed to a poleward displacement of eddy activity (Lu et al., 2008; Zhou et al., 2022). To evaluate the circulation responses in NeuralGCM's uniform-warming simulations, we computed the response of Eliassen-Palm (E-P) flux and diagnosed the convergence of its components to separate the contributions from EMFC ($-\frac{1}{a\cos\phi}\frac{\partial}{\partial\phi}(\overline{u'v'}\cos\phi)$) and eddy heat fluxes convergence (EHFC; $f a\cos\phi \frac{\partial}{\partial p}\left(\frac{\overline{v'\theta'}}{\theta_p}\right)$) (Text S3). In CMIP6 AMIP-P4K, we find enhanced convergence of eddy fluxes at and above the core of the westerly jet (shading in Supplementary Fig. S9a). Both EMFC and EHFC contribute to the strengthened westerly jet (Fig. 8E & Supplementary Fig. S9C). Notably, the enhanced and weakened EMFC straddling the mean westerly jet in the extratropics, which, together with the enhanced convergence above the core of the westerly jet (Fig. 8E), drives the poleward shift and strengthening of the westerlies (Fig. 8A). In NeuralGCM 36-year AMIP-PxK runs, the significantly enhanced EMFC at and above the core of the westerly jet is captured by NeuralGCM. Besides, NeuralGCM reproduces the weakening at the equatorward flank of the westerly jet but fails to capture the enhancement at the poleward flank (Fig. 8F). As a result, NeuralGCM partially reproduces the strengthening of the jet but does not simulate a significant poleward shift (Fig. 8B). In SH, significantly enhanced EMFC 20ºS~30ºS and 150~300 hPa contributes to an equatorward shift of westerly jet in NeuralGCM 36-year AMIP-PxK simulations. Overall, the strengthening of the westerly jet in NeuralGCM is weaker than in CMIP6 AMIP-P4K. A key weakness lies in EHFC. NeuralGCM fails to capture the climatological structure and response pattern. Specifically, it overestimates climatological EHFC at mid-high latitudes in both hemispheres (30º~80º, 200~400 hPa; contours in Supplementary Fig. S9D), owing to weak and excessive meridional eddy heat flux ($\overline{v'T'}$) between 1000 hPa – 500 hPa and 250 hPa – 150 hPa ($\pm15$ K m$^{-1}$ s$^{-1}$; Supplementary Fig. S10B), respectively, compared to CMIP6 AMIP runs ($\pm6$~10 K m$^{-1}$ s$^{-1}$; Supplementary Fig. S10A). Consequently, NeuralGCM simulates spurious vertical convergence of $\overline{v'T'}$ between 200 hPa and 400 hPa in the NH and SH mid-high latitudes (Supplementary Fig. S10D) – a feature absent in both reanalysis (Shaw & Perlwitz, 2014) and CMIP6 AMIP runs (Supplementary Fig. S10C).

The simulations of tropospheric circulation indicate that NeuralGCM falls short in capturing the dynamics of the eddy response or diabatic processes in a moist model. This might raise a requirement to include the fundamental role of precipitation and diabatic heating at midlatitudes, as the diabatic heating in physics-based models dominated by the resolved large-scale precipitation acts as a key driver of midlatitude circulation (e.g., Lachmy, 2022; Ghosh et al., 2024; White et al., 2024; Lubis et al., 2025). As the precipitation simulated by NeuralGCM was highly dependent on the training using the IMERG precipitation dataset (Yuval et al., 2026), whether the ML parameterization scheme could capture the relationship between precipitation and diabatic heating at midlatitudes needs to be further explored in future studies.

The above results indicate that NeuralGCM can capture global-scale responses and some main features of tropospheric large-scale responses under out-of-distribution forcings, consistent with current physics-based ESMs. Despite that, there are pronounced distinctions in the westerly jet shift and responses at UTLS between NeuralGCM and physics-based ESMs, because NeuralGCM might be limited by the training datasets and the lack of $CO_2$ and ozone as input, and hence, falls short in encoding the thermodynamics and dynamics of the middle-atmosphere circulation.

## 4. Conclusions and discussion

ML-based models offer the potential to advance our understanding of the Earth system while reducing computational cost compared to traditional physics-based models. Yet, to establish credibility, such models must undergo systematic evaluation to ensure they can reproduce observed phenomena across a broad range of timescales and generalize to out-of-distribution forcings. In this study, we evaluate the stochastic version of NeuralGCM, a hybrid model combining an atmospheric dynamical core with ML-based component at 2.8º resolution, using a series of idealized experiments spanning synoptic perturbations, interannual SST variability, and out-of-distribution uniform SST warming scenarios. Our key findings are:

(1) NeuralGCM captures the development and evolution of oceanic ETCs. During the first three days after cyclone formation over the North Pacific, NeuralGCM reproduces upper-level PV gradients and associated circulations with skill comparable to observations and physics-based ESMs. Its performance in both forecast-style and long-term AMIP experiments perform comparable to CMIP6 AMIP simulations. A notable weakness is the overestimation of ETC spatial extent, linked to an excessively strong PV gradient over the central-to-eastern North Pacific.

(2) During ENSO events, NeuralGCM reproduces tropical atmospheric responses and the PNA teleconnection. Its performance in simulating anomalous geopotential height at 200 hPa (PCC = 0.86) is comparable to current physics-based models. NeuralGCM also captures aspects of the nonlinear ENSO response, although the simulated anomalies are stronger than those in observations and CMIP6 models. A key limitation is its reduced accuracy in reproducing the magnitude and spatial structure of ENSO-related asymmetries.

(3) NeuralGCM simulates large-scale responses to out-of-distribution uniform SST warmings that are broadly consistent with physics-based ESMs. Its global-mean surface air temperature and precipitation sensitivities (0.86 K K$^{-1}$ and 4.14 % K$^{-1}$, respectively) closely match CMIP6 AMIP-P4K, albeit with some discrepancies. NeuralGCM also reproduces important large-scale circulation features, including amplified warming in the tropical upper-tropospheric amplification and strengthening of the westerly jet. Despite that, significant discrepancies occur in the upper troposphere and lower stratosphere (UTLS). NeuralGCM fails to reproduce warming responses at UTLS, overestimates polar upper-level warming, weakens the meridional temperature gradients, and underrepresents eddy flux convergence, leading to only partial reproduction of the jet strengthening and poleward shifts evident in CMIP6 simulations.

Taken together, these idealized experiments demonstrate that NeuralGCM achieves overall performance comparable to contemporary physics-based ESMs, despite several notable deficiencies. Unlike recent emulators trained directly on idealized $CO_2$-forcing ensemble simulations (e.g., Clark et al., 2024), NeuralGCM was trained exclusively on ERA5 reanalysis (2001–2018), yet it successfully reproduces key aspects of the uniform-warming responses seen in physics-based models. This suggests that integrating a dynamical core with ML techniques enhances the model's generalization beyond the training distributions, although NeuralGCM may still be susceptible to instability under strong forcing. Our analysis is limited to uniform warmings of 1K~4K, covering the range of the projected end-of-century warming across emission scenarios (IPCC, 2021; Lee et al., 2021). Whether current physics-based or ML-based models can generate physically realistic responses under larger warming remains an open question for future studies.

Systematic biases remain. Kochkov et al. (2024) reported that NeuralGCM captures several robust features of warming responses under modest SST increases, while responses diverge under substantial SST warmings. Consistent with this, we find that warming responses under the 2K SST uniform warmings are much closer to the physics-based models than those under the 4K scenario (Figure 7A). More importantly, although NeuralGCM appears to encode tropospheric dynamics reasonably well, it fails to adequately represent processes in the UTLS and middle atmosphere. Similar deficiencies were also noted by Liang et al. (2025) when forcing NeuralGCM with decreasing sea-ice concentrations. These shortcomings may be related to limitations in vertical resolution and inadequate learning of stratosphere dynamics (Baxter et al., 2025). Although 11 of NeuralGCM's 37 vertical levels lie above 100 hPa, its loss function only extends to 30 hPa, and only four stratospheric levels (100, 70, 50, 30 hPa) are directly constrained during training (Kochkov et al., 2024). This limitation likely contributes to biases in the UTLS mean state and response and may also impair simulation of key modes of climate variability (Baxter et al., 2025). Therefore, the training process and model structure might directly lead to the biases in mean states and responses.

Additionally, NeuralGCM exhibits errors in tropical convection, westerly jet shifts, and stratospheric circulation – likely arising from imperfect linkages between diabatic heating and large-scale dynamics, as well as limitations in the representation of vertical heat flux. These deficiencies hinder its ability to fully capture the warming responses of the Hadley cell's ascending branch and in stratospheric temperature and circulation. Improvements in convection parameterization, vertical energy transport, and large-scale circulation representation are therefore critical for enhancing model fidelity. Finally, the horizontal and vertical resolutions used in this study are much coarser than those of many current ESMs. This coarse resolution may dampen convective processes, while contributing to overestimation of ETC size and PV gradients, and an underestimation of equatorial convection responses to uniform SST warmings. A previous work has emphasized that ML-based models must balance forecasting skill with physical realism (Bonavita, 2024). Embedding robust physical constraints – such as dynamical cores – within ML frameworks, along with increasing the spatial and vertical resolution, offers a promising pathway to mitigate these biases and improve generalization.

In summary, we proposed and conducted a suite of idealized evaluation experiments for ML-based models, consistent with the recommended "Menu for ML-based ESM Evaluation" (Ullrich et al., 2025). These experiments provide a structured framework for benchmarking hybrid and ML-based climate models against both observations and physics-based ESMs. Future work should expand this framework to include additional Earth system components, providing a foundation for a comprehensive, systematic evaluation of next-generation ML-based models.

**Acknowledgments**

We thank the World Climate Research Programme's Working Group on Coupled Modeling, which is responsible for CMIP6, and the climate modeling groups. This study is supported by Office of Science, U.S. Department of Energy (DOE) Biological and Environmental Research as part of the Regional and Global Model Analysis program area through the Water Cycle: Modeling of Circulation, Convection, and Earth System Mechanisms (WACCEM) scientific focus area. The Pacific Northwest National Laboratory (PNNL) is operated for DOE by Battelle Memorial Institute under contract DE-AC05-76RL01830. GZ is supported by U.S. NSF awards AGS-2327959 and RISE-2530555. JL was also supported by the startup projects of Ocean University of China: project 30010000-862401013230 and project 3001000-862505020010. We also acknowledge Prof. Zhuo Wang for the help and suggestions on the estimation of NeuralGCM warming responses.

**Author contributions:** Z.C. and L.R.L. designed the research. Z.C. conducted the simulations and wrote the draft. W.Z., J.L., Y.L., C.C.C., S.L., and B.H. contributed the diagnostic analysis. L.R.L., Y.W., G.Z., and Y.Q. contributed to improving the analysis and interpretation. All authors edited the paper.

**Competing Interests:** The authors declare no competing interests.

**Data Availability Statement:** All data needed to evaluate the conclusions in the paper are presented in the paper, the Supplementary Materials, and/or the linked repositories. The CMIP6 data are acquired from https://aims2.llnl.gov/search/cmip6/. The NeuralGCM simulations are available from

https://portal.nersc.gov/cfs/m1867/zmchen/Work/2024/NGCM_Performance/Data/.

**References**

Acosta, M. C., Palomas, S., Paronuzzi Ticco, S. V., Utrera, G., Biercamp, J., Bretonniere, P. A., Budich, R., Castrillo, M., Caubel, A., Doblas-Reyes, F., Epicoco, I., Fladrich, U., Joussaume, S., Kumar Gupta, A., Lawrence, B., Le Sager, P., Lister, G., Moine, M. P., Rioual, J. C., … Balaji, V. (2024). The computational and energy cost of simulation and storage for climate science: lessons from CMIP6. *Geoscientific Model Development*, *17*(8), 3081–3098. https://doi.org/10.5194/gmd-17-3081-2024

Adam, O., Schneider, T., Brient, F., & Bischoff, T. (2016). Relation of the double-ITCZ bias to the atmospheric energy budget in climate models. *Geophysical Research Letters*, *43*(14), 7670–7677. https://doi.org/10.1002/2016GL069465

Adler, R. F., Huffman, G. J., Chang, A., Ferraro, R., Xie, P. P., Janowiak, J., Rudolf, B., Schneider, U., Curtis, S., Bolvin, D., Gruber, A., Susskind, J., Arkin, P., & Nelkin, E. (2003). The version-2 global precipitation climatology project (GPCP) monthly precipitation analysis (1979-present). *Journal of Hydrometeorology*, *4*(6), 1147–1167. https://doi.org/10.1175/1525-7541(2003)004<1147:Tvgpcp>2.0.Co;2

Andrews, D. G., Holton, J. R., & Leovy, B. C. (1987). *Middle atmosphere dynamics* (R. Domowska & J. R. Holton (eds.); Vol. 40, Issue 2). Academic Press. https://doi.org/10.1016/0019-1035(88)90078-4

Aquila, V., Swartz, W. H., Waugh, D. W., Colarco, P. R., Pawson, S., Polvani, L. M., & Stolarski, R. S. (2016). Isolating the roles of different forcing agents in global stratospheric temperature changes using model integrations with incrementally added single forcings. *Journal of Geophysical Research: Atmosphere*, *121*, 8067–8082. https://doi.org/10.1002/2015JD023841

Baño-Medina, J., Sengupta, A., Doyle, J. D., Reynolds, C. A., Watson-Parris, D., & Monache, L. D. (2025). Are AI weather models learning atmospheric physics? A sensitivity analysis of cyclone Xynthia. *Npj Climate and Atmospheric Science*, *8*(1). https://doi.org/10.1038/s41612-025-00949-6

Baxter, I., Pahlavan, H., Hassanzadeh, P., Rucker, K., & Shaw, T. (2025). *Benchmarking atmospheric circulation variability in an AI emulator, ACE2, and a hybrid model, NeuralGCM*. 1–25. http://arxiv.org/abs/2510.04466

Bi, K., Xie, L., Zhang, H., Chen, X., Gu, X., & Tian, Q. (2023). Accurate medium-range global

weather forecasting with 3D neural networks. *Nature*, *619*(7970), 533–538. https://doi.org/10.1038/s41586-023-06185-3

Bonavita, M. (2024). On Some Limitations of Current Machine Learning Weather Prediction Models. *Geophysical Research Letters*, *51*(12). https://doi.org/10.1029/2023GL107377

Bouallègue, Z. Ben, Clare, M. C. A., Magnusson, L., Gascón, E., Maier-Gerber, M., Janoušek, M., Rodwell, M., Pinault, F., Dramsch, J. S., Lang, S. T. K., Raoult, B., Rabier, F., Chevallier, M., Sandu, I., Dueben, P., Chantry, M., & Pappenberger, F. (2024). The Rise of Data-Driven Weather Forecasting A First Statistical Assessment of Machine Learning–Based Weather Forecasts in an Operational-Like Context. *Bulletin of the American Meteorological Society*, *105*(6), E864–E883. https://doi.org/10.1175/BAMS-D-23-0162.1

Brenowitz, N. D., Cohen, Y., Pathak, J., Mahesh, A., Bonev, B., Kurth, T., Durran, D., Harrington, P., & Pritchard, M. S. (2024). *A Practical Probabilistic Benchmark for AI Weather Models*. 1–16.

Bulgin, C. E., Merchant, C. J., & Ferreira, D. (2020). Tendencies, variability and persistence of sea surface temperature anomalies. *Scientific Reports*, *10*(1), 1–13. https://doi.org/10.1038/s41598-020-64785-9

Butchart, N., & Butchart, N. (2014). The Brewer-Dobson circulation. *Rev. Geophys*, *52*, 157–184. https://doi.org/10.1002/2013RG000448.One

Camps-Valls, G., Fernández-Torres, M. Á., Cohrs, K. H., Höhl, A., Castelletti, A., Pacal, A., Robin, C., Martinuzzi, F., Papoutsis, I., Prapas, I., Pérez-Aracil, J., Weigel, K., Gonzalez-Calabuig, M., Reichstein, M., Rabel, M., Giuliani, M., Mahecha, M. D., Popescu, O. I., Pellicer-Valero, O. J., … Williams, T. (2025). Artificial intelligence for modeling and understanding extreme weather and climate events. *Nature Communications* , *16*(1). https://doi.org/10.1038/s41467-025-56573-8

Chemke, R., & Coumou, D. (2024). Human in fl uence on the recent weakening of storm tracks in boreal summer. *Npj Climate and Atmospheric Science*, *7*(86), 1–8. https://doi.org/10.1038/s41612-024-00640-2

Chemke, R., Ming, Y., & Yuval, J. (2022). The intensification of winter mid-latitude storm tracks in the Southern Hemisphere Rei. *Nature Climate Change*, *12*(June), 553–557. https://doi.org/10.1038/s41558-022-01368-8

Chemke, R., & Yuval, J. (2023). Human-induced weakening of the Northern Hemisphere

tropical circulation. *Nature*, *617*, 529–532. https://doi.org/10.1038/s41586-023-05903-1

Chen, D., Dai, A., & Hall, A. (2021). The Convective-To-Total Precipitation Ratio and the "Drizzling" Bias in Climate Models. *Journal of Geophysical Research: Atmospheres*, *126*(16), 1–17. https://doi.org/10.1029/2020JD034198

Chen, Z., Zhou, T., Chen, X., Zhang, L., Qian, Y., Wang, Z., He, L., & Leung, L. R. (2024). Understanding the biases in global monsoon simulations from the perspective of atmospheric energy transport. *Journal of Climate*, 4647–4666. https://doi.org/10.1175/jcli-d-23-0444.1

Christopoulos, C., & Schneider, T. (2021). Assessing Biases and Climate Implications of the Diurnal Precipitation Cycle in Climate Models. *Geophysical Research Letters*, *48*(13), 1–9. https://doi.org/10.1029/2021GL093017

Clark, S. K., Watt-Meyer, O., Kwa, A., McGibbon, J., Henn, B., Perkins, W. A., Wu, E., Harris, L. M., & Bretherton, C. S. (2024). *ACE2-SOM: Coupling an ML atmospheric emulator to a slab ocean and learning the sensitivity of climate to changed CO$_2$*. 1–25. http://arxiv.org/abs/2412.04418

Diao, C., & Barnes, E. A. (2025). *Assessing MJO Tropical-Extratropical Teleconnections in Deep Learning Weather Prediction Models Key Points : scientific investigations and dynamical hypothesis testing Assessing MJO Tropical - Extratropical Teleconnections in Deep Learning Weather Predict*.

Duan, S., Zhang, J., Bonfils, C., & Pallotta, G. (2025). Testing NeuralGCM's capability to simulate future heatwaves based on the 2021 Paci c Northwest heatwave event. *Npj Climate and Atmospheric Science*, *8*(251), 1–8. https://doi.org/10.1038/s41612-025-01137-2

Eyring, V., Bony, S., Meehl, G. A., Senior, C. A., Stevens, B., Stouffer, R. J., & Taylor, K. E. (2016). Overview of the Coupled Model Intercomparison Project Phase 6 (CMIP6) experimental design and organization. *Geoscientific Model Development*, *9*(5), 1937–1958. https://doi.org/10.5194/gmd-9-1937-2016

Eyring, V., Collins, W. D., Gentine, P., Barnes, E. A., Barreiro, M., Beucler, T., Bocquet, M., Bretherton, C. S., Christensen, H. M., Dagon, K., Gagne, D. J., Hall, D., Hammerling, D., Hoyer, S., Iglesias-Suarez, F., Lopez-Gomez, I., McGraw, M. C., Meehl, G. A., Molina, M. J., … Zanna, L. (2024b). Pushing the frontiers in climate modelling and analysis with machine learning. *Nature Climate Change*, *14*(9), 916–928. https://doi.org/10.1038/s41558-

024-02095-y

Eyring, V., Gentine, P., Camps-Valls, G., Lawrence, D. M., & Reichstein, M. (2024a). AI-empowered next-generation multiscale climate modelling for mitigation and adaptation. *Nature Geoscience*. https://doi.org/10.1038/s41561-024-01527-w

Flato, G., Marotzke, J., Abiodun, B., Braconnot, P., Chou, S. C., Collins, W., Cox, P., Driouech, F., Emori, S., Eyring, V., Forest, C., Gleckler, P., Guilyardi, E., Jakob, C., Kattsov, V., Reason, C., & Rummukainen, M. (2013). Evaluation of climate models. In T. F. Stocker, D. Qin, G.-K. Plattner, M. Tignor, S. K. Allen, J. Boschung, A. Nauels, & Y. Xia (Eds.), *Climate Change 2013 the Physical Science Basis: Working Group I Contribution to the Fifth Assessment Report of the Intergovernmental Panel on Climate Change* (Vol. 9781107057, pp. 741–866). Cambridge University Press. https://doi.org/10.1017/CBO9781107415324.020

Frauen, C., Dommenget, D., Tyrrell, N., Rezny, M., & Wales, S. (2014). Analysis of the nonlinearity of El Niño-Southern Oscillation teleconnections. *Journal of Climate*, *27*(16), 6225–6244. https://doi.org/10.1175/JCLI-D-13-00757.1

Garfinkel, C. I., Weinberger, I., White, I. P., Oman, L. D., Aquila, V., & Lim, Y. K. (2019). The salience of nonlinearities in the boreal winter response to ENSO: North Pacific and North America. *Climate Dynamics*, *52*(7–8), 4429–4446. https://doi.org/10.1007/s00382-018-4386-x

Ghosh, S., Lachmy, O., & Kaspi, Y. (2024). The Role of Diabatic Heating in the Midlatitude Atmospheric Circulation Response to Climate Change. *Journal of Climate*, *37*(10), 2987–3009. https://doi.org/10.1175/JCLI-D-23-0345.1

Gill, A. E. (1980). Some simple solutions for heat-induced tropical circulation. *Quarterly Journal of the Royal Meteorological Socitey*, *106*(449), 447–462. https://doi.org/10.1002/qj.49710644905

Gyakum, J. R., & Danielson, R. E. (2000). Analysis of meteorological precursors to ordinary and explosive cyclogenesis in the Western North Pacific. *Monthly Weather Review*, *128*(3), 851–863. https://doi.org/10.1175/1520-0493(2000)128<0851:AOMPTO>2.0.CO;2

Hakim, G. J. (2003). Developing wave packets in the North Pacific storm track. *Monthly Weather Review*, *131*(11), 2824–2837. https://doi.org/10.1175/1520-0493(2003)131<2824:DWPITN>2.0.CO;2

Hakim, G. J., & Masanam, S. (2024). Dynamical Tests of a Deep Learning Weather Prediction Model. *Artificial Intelligence for the Earth Systems*, *3*(3), 1–11. https://doi.org/10.1175/aies-d-23-0090.1

Held, I. M. (1993). Large-Scale Dynamics and Global Warming OF ARTICLES REPORTING ON THE U.S . AND THE HUMAN DIMENSIONS OF GLOBAL GLOBAL CHANGE , BUT DO NOT THE NATIONAL ACADEMY OF SCIENCES ,. *Bulletin of the American Meteorological Society*, *74*(2), 228–242.

Heo, K. Y., Ha, K. J., & Ha, T. (2019). Explosive cyclogenesis around the Korean Peninsula in May 2016 from a potential vorticity perspective: Case study and numerical simulations. *Atmosphere*, *10*(6). https://doi.org/10.3390/atmos10060322

Hersbach, H., Bell, W., Berrisford, P., Horányi, A., M-S, J., Nicolas, J., Radu, R., Schepers, D., Simmons, A., Soci, C., & Dee, D. (2019). Global reanalysis: goodbye ERA-Interim, hello ERA5. *Meteorology*, *159*, 17–24. https://doi.org/10.21957/vf291hehd7

Huang, B., Thorne, P. W., Banzon, V. F., Boyer, T., Chepurin, G., Lawrimore, J. H., Menne, M. J., Smith, T. M., Vose, R. S., & Zhang, H. M. (2017). Extended reconstructed Sea surface temperature, Version 5 (ERSSTv5): Upgrades, validations, and intercomparisons. *Journal of Climate*, *30*(20), 8179–8205. https://doi.org/10.1175/JCLI-D-16-0836.1

Husain, S. Z., Separovic, L., Caron, J.-F., Aider, R., Buehner, M., Chamberland, S., Lapalme, E., McTaggart-Cowan, R., Subich, C., Vaillancourt, P. A., Yang, J., & Zadra, A. (2025). Leveraging data-driven weather models for improving numerical weather prediction skill through large-scale spectral nudging. *Weather and Forecasting*, 1–48. https://doi.org/10.1175/waf-d-24-0139.1

IPCC. (2021). Summary for Policymakers. In V. Masson-Delmotte, P. Zhai, A. Pirani, S. L. Connors, C. Péan, S. Berger, N. Caud, Y. Chen, L. Goldfarb, M. I. Gomis, M. Huang, K. Leitzell, E. Lonnoy, J. B. R. Matthews, T. K. Maycock, T. Waterfield, O. Yelekçi, R. Yu, & B. Zhou (Eds.), *In: Climate Change 2021: The Physical Science Basis. Contribution of Working Group I to the Sixth Assessment Report of the Intergovernmental Panel on Climate Changeental Panel on Climate Change* (p. 42). Cambridge University Press. https://www.ipcc.ch/report/ar6/wg1/

Jiménez-Esteve, B., Barriopedro, D., Johnson, J. E., & García-Herrera, R. (2025). AI-Driven Weather Forecasts to Accelerate Climate Change Attribution of Heatwaves. *Earth's Future*,

13(8), e2025EF006453. https://doi.org/10.1029/2025EF006453

Jiménez-Esteve, B., & Domeisen, D. I. V. (2019). Nonlinearity in the North Pacific Atmospheric Response to a Linear ENSO Forcing. *Geophysical Research Letters*, *46*(4), 2271–2281. https://doi.org/10.1029/2018GL081226

Johnson, N. C., & Xie, S.-P. (2010). Changes in the sea surface temperature threshold for tropical convection. *Nature Geoscience*, *3*(12), 842–845. https://doi.org/10.1038/ngeo1008

Kautz, L. A., Martius, O., Pfahl, S., Pinto, J. G., Ramos, A. M., Sousa, P. M., & Woollings, T. (2022). Atmospheric blocking and weather extremes over the Euro-Atlantic sector - A review. *Weather and Climate Dynamics*, *3*(1), 305–336. https://doi.org/10.5194/wcd-3-305-2022

Keisler, R. (2022). *Forecasting Global Weather with Graph Neural Networks*. 1–16. http://arxiv.org/abs/2202.07575

Kent, C., Scaife, A. A., Dunstone, N. J., Smith, D., Hardiman, S. C., & Watt-meyer, O. (2025). Skilful global seasonal predictions from a machine learning weather model trained on reanalysis data. *Npj Climate and Atmospheric Science*, *8*(314), 1–17. https://doi.org/10.1038/s41612-025-01198-3

Kochkov, D., Yuval, J., Langmore, I., Norgaard, P., Smith, J., Mooers, G., Klöwer, M., Lottes, J., Rasp, S., Düben, P., Hatfield, S., Battaglia, P., Sanchez-Gonzalez, A., Willson, M., Brenner, M. P., & Hoyer, S. (2024). Neural general circulation models for weather and climate. *Nature*, *November 2023*. https://doi.org/10.1038/s41586-024-07744-y

Lachmy, O. (2022). The Relation Between the Latitudinal Shifts of Midlatitude Diabatic Heating, Eddy Heat Flux, and the Eddy-Driven Jet in CMIP6 Models. *Journal of Geophysical Research: Atmospheres*, *127*(16), 1–19. https://doi.org/10.1029/2022JD036556

Lam, R., Sanchez-Gonzalez, A., Willson, M., Wirnsberger, P., Fortunato, M., Alet, F., Ravuri, S., Ewalds, T., Eaton-Rosen, Z., Hu, W., Merose, A., Hoyer, S., Holland, G., Vinyals, O., Stott, J., Pritzel, A., Mohamed, S., & Battaglia, P. (2023). Learning skillful medium-range global weather forecasting. *Science*, *382*(6677), 1416–1422. https://doi.org/10.1126/science.adi2336

Lee, J. Y., Marotzke, J., Bala, G., Cao, L., Corti, S., Dunne, J. P., Engelbrecht, F., Fischer, E., Fyfe, J. C., Jones, C., Maycock, A., Mutemi, J., Ndiaye, O., Panickal, S., & Zhou, T. (2021). Chapter 4: Future global climate: scenario-based projections and near-term42 information. In V. Masson-Delmotte, P. Zhai, A. Pirani, S. L. Connors, C. Péan, S. Berger, N. Caud, Y. Chen, L. Goldfarb, M. I. Gomis, M. Huang, K. Leitzell, E. Lonnoy, J. B. R. Matthews, T. K. Maycock, T. Waterfield, O. Yelekçi, R. Yu, & B. Zhou (Eds.), *Climate Change 2021: The Physical Science Basis. Contribution of Working Group I to the Sixth Assessment Report of the Intergovernmental Panel on Climate Change*. Cambridge University Press.

Li, G., & Xie, S. P. (2012). Origins of tropical-wide SST biases in CMIP multi-model ensembles. *Geophysical Research Letters*, *39*(22), 1–5. https://doi.org/10.1029/2012GL053777

Liang, Y. C., Lutsko, N. J., & Kwon, Y. O. (2025). *Exploring the Atmospheric Responses to Arctic Sea - Ice Loss in Google ' s NeuralGCM*. https://doi.org/10.1029/2025MS005264

Lu, J., Chen, G., & Frierson, D. M. W. (2008). Response of the zonal mean atmospheric circulation to El Niño versus global warming. *Journal of Climate*, *21*(22), 5835–5851. https://doi.org/10.1175/2008JCLI2200.1

Lu, J., Chen, G., Leung, L. R., Burrows, D. A., Yang, Q., Sakaguchi, K., & Hagos, S. (2015). Toward the dynamical convergence on the jet stream in aquaplanet AGCMs. *Journal of Climate*, *28*(17), 6763–6782. https://doi.org/10.1175/JCLI-D-14-00761.1

Lubis, S. W., Harrop, B. E., Lu, J., Leung, L. R., Chen, Z., Huang, C. S. Y., & Omrani, N. (2025). Cloud radiative effects signi fi cantly increase wintertime atmospheric blocking in the Euro-Atlantic sector. *Nature Communications*, *16*(9763), 1–17. https://doi.org/10.1038/s41467-025-64672-9

Lubis, S. W., Mattes, K., Harnik, N., Omrani, N. E., & Wahl, S. (2018). Downward Wave Coupling between the Stratosphere and Troposphere under Future Anthropogenic Climate Change. *Journal of Climate*, *31*(10), 4135–4155. https://doi.org/10.1175/JCLI-D-17-0382.1

Lubis, S. W., Omrani, N. E., Matthes, K., & Wahl, S. (2016). Impact of the Antarctic ozone hole on the vertical coupling of the stratosphere-mesosphere-lower thermosphere system. *Journal of the Atmospheric Sciences*, *73*(6), 2509–2528. https://doi.org/10.1175/JAS-D-15-0189.1

Matsuno, T. (1966). Quasi-Geostrophic Motions in the Equatorial Area. *Journal of the Meteorological Society of Japan. Ser. II*, *44*(1), 25–43. https://doi.org/10.2151/jmsj1965.44.1_25

Meehl, G. A., Senior, C. A., Eyring, V., Gregory, F., Lamarque, J.-F., Stouffer, R. J., Taylor, K. E., & Schlund, M. (2020). Context for interpreting equilibrium climate sensitivity and transient climate response from the CMIP6 Earth system models. *Science Advances*, *6*, eaba1981. https://doi.org/10.1126/sciadv.aba1981

Mitchell, D. M. (2016). Attributing the forced components of observed stratospheric temperature variability to external drivers. *Quarterly Journal of the Royal Meteorological Society*, *142*, 1041–1047. https://doi.org/10.1002/qj.2707

Ni, J., Fu, G., Chen, L. J., & Li, P. Y. (2025). Perspective of explosive cyclones with three shapes of upper-level potential vorticity over the northern Atlantic Ocean: Perspective of explosive cyclones with three shapes of upper-level potential vorticity: J. Ni et al. *Climate Dynamics*, *63*(1). https://doi.org/10.1007/s00382-024-07482-x

Palipane, E., Lu, J., Staten, P., Chen, G., & Schneider, E. K. (2017). Investigating the zonal wind response to SST warming using transient ensemble AGCM experiments. *Climate Dynamics*, *48*(1–2), 523–540. https://doi.org/10.1007/s00382-016-3092-9

Pathak, J., Cohen, Y., Garg, P., Harrington, P., Brenowitz, N., Durran, D., Mardani, M., Vahdat, A., Xu, S., Kashinath, K., & Pritchard, M. (2024). *Kilometer-Scale Convection Allowing Model Emulation using Generative Diffusion Modeling*. 1–33. http://arxiv.org/abs/2408.10958

Peings, Y., Dong, C., Mahesh, A., Pritchard, M. S., Collins, W. D., & Magnusdottir, G. (2025). *Subseasonal forecasting and MJO teleconnections in machine learning weather prediction models*. https://doi.org/10.22541/au.175207564.41006123/v1

Pithan, F., Shepherd, T. G., Zappa, G., & Sandu, I. (2016). Climate model biases in jet streams, blocking and storm tracks resulting from missing orographic drag. *Geophysical Research Letters*, *43*(13), 7231–7240. https://doi.org/10.1002/2016GL069551

Price, I., Sanchez-Gonzalez, A., Alet, F., Andersson, T. R., El-Kadi, A., Masters, D., Ewalds, T., Stott, J., Mohamed, S., Battaglia, P., Lam, R., & Willson, M. (2024). Probabilistic weather forecasting with machine learning. *Nature*, *April*, 8–11. https://doi.org/10.1038/s41586-024-08252-9

Rasp, S., Hoyer, S., Merose, A., Langmore, I., Battaglia, P., Russell, T., Sanchez-Gonzalez, A., Yang, V., Carver, R., Agrawal, S., Chantry, M., Ben Bouallegue, Z., Dueben, P., Bromberg, C., Sisk, J., Barrington, L., Bell, A., & Sha, F. (2024). WeatherBench 2: A Benchmark for the Next Generation of Data-Driven Global Weather Models. *Journal of Advances in Modeling Earth Systems*, *16*(6). https://doi.org/10.1029/2023MS004019

Rayner, N. A., Parker, D. E., Horton, E. B., Folland, C. K., Alexander, L. V, Rowell, D. P., Kent, E. C., & Kaplan, A. (2003). Global analyses of sea surface temperature, sea ice, and night marine air temperature since the late nineteenth century. *Journal of Geophysical Research*, *108*(D14). https://doi.org/10.1029/2002JD002670

Ren, Z., & Zhou, T. (2024). Understanding the alleviation of "Double-ITCZ" bias in CMIP6 models from the perspective of atmospheric energy balance. *Climate Dynamics*, *62*(7), 5769–5786. https://doi.org/10.1007/s00382-024-07238-7

Rosenlof, K. H. (1995). Seasonal cycle of the residual mean meridional circulation in the stratosphere. *Journal of Geophysical Research-Atmospheres*, *100*, 5173–5191. https://doi.org/10.1029/94jd03122

Seneviratne, S. I., Zhang, X., Adnan, M., Badi, W., Dereczynski, C., Di Luca, A., Ghosh, S., Iskandar, I., Kossin, J., Lewis, S., Otto, F., Pinto, I., Satoh, M., M., V.-S. S., Wehner, M., & Zhou, B. (2021). Weather and Climate Extreme Events in a Changing Climate Supplementary Material. In V. Masson-Delmotte, P. Zhai, A. Pirani, S. L. Connors, C. Péan, S. Berger, N. Caud, Y. Chen, L. Goldfarb, M. I. Gomis, M. Huang, K. Leitzell, E. Lonnoy, J. B. R. Matthews, T. K. Maycock, T. Waterfield, O. Yelekçi, R. Yu, & B. Zhou (Eds.), *Climate Change 2021: The Physical Science Basis. Contribution of Working Group I to the Sixth Assessment Report of the Intergovernmental Panel on Climate Change* (pp. 1–14). Cambridge University Press. chrome-extension://efaidnbmnnnibpcajpcglclefindmkaj/https://www.ipcc.ch/report/ar6/wg1/downloads/report/IPCC_AR6_WGI_Chapter11_SM.pdf

Shaw, T. A., & Perlwitz, J. (2014). On the control of the residual circulation and stratospheric temperatures in the arctic by planetary wave coupling. *Journal of the Atmospheric Sciences*, *71*(1), 195–206. https://doi.org/10.1175/JAS-D-13-0138.1

Shine, B. K. P., Bourqui, M. S., Forster, P. M. D. F., Hare, S. H. E., & Langematz, U. (2003). A comparison of model-simulated trends in stratospheric temperatures. *Quarterly Journal of the Royal Meteorological Society*, *129*, 1565–1588. https://doi.org/10.1256/qj.02.186

Sun, Y. Q., Hassanzadeh, P., Zand, M., Chattopadhyay, A., Weare, J., & Abbot, D. S. (2025). Can AI weather models predict out-of-distribution gray swan tropical cyclones?

*Proceedings of the National Academy of Sciences of the United States of America*, *122*(21), 1–10. https://doi.org/10.1073/pnas.2420914122

Tang, L., Lu, R., Lin, Z., Lu, J., & Chen, Z. (2024). Concurrent Inter-Model Spread of Boreal Winter Westerly Jet Meridional Positions Between the Northern and Southern Hemispheres in CMIP6 Models. *International Journal of Climatology*, 5474–5486. https://doi.org/10.1002/joc.8647

Tebaldi, C., Debeire, K., Eyring, V., Fischer, E., Fyfe, J., Friedlingstein, P., Knutti, R., Lowe, J., O'Neill, B., Sanderson, B., Van Vuuren, D., Riahi, K., Meinshausen, M., Nicholls, Z., Tokarska, K., Hurtt, G., Kriegler, E., Meehl, G., Moss, R., … Ziehn, T. (2021). Climate model projections from the Scenario Model Intercomparison Project (ScenarioMIP) of CMIP6. *Earth System Dynamics*, *12*(1), 253–293. https://doi.org/10.5194/esd-12-253-2021

Tian, B., & Dong, X. (2020). The Double-ITCZ Bias in CMIP3, CMIP5, and CMIP6 Models Based on Annual Mean Precipitation. *Geophysical Research Letters*, *47*(8), 1–11. https://doi.org/10.1029/2020GL087232

Ullrich, P. A., Barnes, E. A., Collins, W. D., Dagon, K., Duan, S., Elms, J., Lee, J., Leung, L. R., Lu, D., Molina, M. J., O'Brien, T. A., & Rebassoo, F. O. (2025). Recommendations for Comprehensive and Independent Evaluation of Machine Learning-Based Earth System Models. *Journal of Geophysical Research: Machine Learning and Computation*, *2*(1). https://doi.org/10.1029/2024JH000496

Wang, C., Pritchard, M. S., Brenowitz, N., Cohen, Y., Bonev, B., Kurth, T., Durran, D., & Pathak, J. (2024). *Coupled Ocean-Atmosphere Dynamics in a Machine Learning Earth System Model*. 1–24. http://arxiv.org/abs/2406.08632

Wang, C., Zhang, L., Lee, S.-K., Wu, L., & Mechoso, C. R. (2014). A global perspective on CMIP5 climate model biases. *Nature Climate Change*, *4*(3), 201–205. https://doi.org/10.1038/nclimate2118

Wang, Y., Hu, K., Huang, G., & Tao, W. (2023). The Role of Nonlinear Energy Advection in Forming Asymmetric Structure of ENSO Teleconnections Over the North Pacific and North America. *Geophysical Research Letters*, *50*(17). https://doi.org/10.1029/2023GL105277

Watt-Meyer, O., Dresdner, G., McGibbon, J., Clark, S. K., Henn, B., Duncan, J., Brenowitz, N. D., Kashinath, K., Pritchard, M. S., Bonev, B., Peters, M. E., & Bretherton, C. S. (2023). *ACE: A fast, skillful learned global atmospheric model for climate prediction*.

http://arxiv.org/abs/2310.02074

Webb, M. J., Andrews, T., Bodas-Salcedo, A., Bony, S., Bretherton, C. S., Chadwick, R., Chepfer, H., Douville, H., Good, P., Kay, J. E., Klein, S. A., Marchand, R., Medeiros, B., Siebesma, A. P., Skinner, C. B., Stevens, B., Tselioudis, G., Tsushima, Y., & Watanabe, M. (2017). The Cloud Feedback Model Intercomparison Project (CFMIP) contribution to CMIP6. *Geoscientific Model Development*, *10*(1), 359–384. https://doi.org/10.5194/gmd-10-359-2017

White, I. P., Lachmy, O., & Harnik, N. (2024). Influence of a local diabatic heating source on the midlatitude circulation. *Quarterly Journal of the Royal Meteorological Society*, *150*(765), 5167–5187. https://doi.org/10.1002/qj.4863

Woollings, T., Barriopedro, D., Methven, J., Son, S. W., Martius, O., Harvey, B., Sillmann, J., Lupo, A. R., & Seneviratne, S. (2018). Blocking and its Response to Climate Change. *Current Climate Change Reports*, *4*(3), 287–300. https://doi.org/10.1007/s40641-018-0108-z

Wu, Y., Seager, R., Ting, M., Naik, N., & Shaw, T. A. (2012). Atmospheric circulation response to an instantaneous doubling of carbon dioxide. Part I: Model experiments and transient thermal response in the troposphere. *Journal of Climate*, *25*(8), 2862–2879. https://doi.org/10.1175/JCLI-D-11-00284.1

Yoshida, A., & Asuma, Y. (2004). Structures and environment of explosively developing extratropical cyclones in the northwestern Pacific region. *Monthly Weather Review*, *132*(5), 1121–1142. https://doi.org/10.1175/1520-0493(2004)132<1121:SAEOED>2.0.CO;2

Yuval, J., Langmore, I., Kochkov, D., & Hoyer, S. (2026). Neural general circulation models optimized to predict satellite-based precipitation observations. *Science Advances*, *12*(eadv6891), 1–11. https://doi.org/10.1126/sciadv.adv6891

Zelinka, M. D., Myers, T. A., McCoy, D. T., Po-Chedley, S., Caldwell, P. M., Ceppi, P., Klein, S. A., & Taylor, K. E. (2020). Causes of Higher Climate Sensitivity in CMIP6 Models. *Geophysical Research Letters*, *47*(1). https://doi.org/10.1029/2019gl085782

Zhang, B., & Merlis, T. M. (2025). *The Equilibrium Response of Atmospheric Machine-Learning Models to Uniform Sea Surface Temperature Warming*. 1–23. http://axiv/2510.02415v1

Zhang, C. (1993). Large-scale variability of atmospheric deep convection in relation to sea surface temperature in the tropics. *Journal of Climate*, *6*(10), 1898–1913.

https://doi.org/10.1175/1520-0442(1993)006<1898:LSVOAD>2.0.CO;2

Zhang, G., Rao, M., Yuval, J., & Zhao, M. (2025). Advancing seasonal prediction of tropical cyclone activity with a hybrid AI-physics climate model. *Environmental Research Letters*, *20*(094031), 1–11. https://doi.org/10.1088/1748-9326/adf864

Zhang, Q., Liu, B., Li, S., & Zhou, T. (2023). Understanding Models' Global Sea Surface Temperature Bias in Mean State: From CMIP5 to CMIP6. *Geophysical Research Letters*, *50*(4), 1–11. https://doi.org/10.1029/2022GL100888

Zhang, T., Perlwitz, J., & Hoerling, M. (2014). What is responsible for the strong observed asymmetry in teleconnections between El Niño and La Niña. Geophysical Research Letters, *41*, 1019–1025. https://doi.org/10.1002/2013GL058964

Zhou, C., Chen, L., Zhong, X., Lu, B., Li, H., & Wu, L. (2025). *A machine learning model for skillful climate system prediction*. 1–39. https://doi.org/10.48550/arXiv.2505.06269

Zhou, T., Chen, Z., Zou, L., Chen, X., Yu, Y., Wang, B., Bao, Q., Bao, Y., Cao, J., He, B., Hu, S., Li, L., Li, J., Lin, Y., Ma, L., Qiao, F., Rong, X., Song, Z., Tang, Y., … Zhang, M. (2020). Development of Climate and Earth System Models in China: Past Achievements and New CMIP6 Results. *Journal of Meteorological Research*, *34*(1), 1–19. https://doi.org/10.1007/s13351-020-9164-0

Zhou, W., Leung, L. R., & Lu, J. (2022). Seasonally and regionally dependent shifts of the atmospheric westerly jets under global warming. *Journal of Climate*, *35*(16), 5433–5447. https://doi.org/10.1175/JCLI-D-21-0723.1

Zhou, W., Ruby Leung, L., & Lu, J. (2022). Linking Large-Scale Double-ITCZ Bias to Local-Scale Drizzling Bias in Climate Models. *Journal of Climate*, *35*(24), 4365–4379. https://doi.org/10.1175/JCLI-D-22-0336.1

Zhou, W., Xie, S. P., & Yang, D. (2019). Enhanced equatorial warming causes deep-tropical contraction and subtropical monsoon shift. *Nature Climate Change*, *9*(11), 834–839. https://doi.org/10.1038/s41558-019-0603-9
48